\documentclass[12pt]{article}
\pdfoutput=1
\usepackage{jheppub,amsmath,amsfonts,amsbsy,latexsym,jheppub,amssymb,amscd,amstext}
\usepackage{overpic,youngtab}
\usepackage{array}
\usepackage[scr=euler,cal=cm]{mathalfa}
\usepackage{color}

\def\be{\begin{equation}}
\def\ee{\end{equation}}
\def\bea#1\eea{\begin{align}#1\end{align}}
\def\pd{\partial}
\def\a{\alpha}
\def\b{\beta}
\def\g{\gamma}
\def\d{\delta}
\def\m{\mu}
\def\n{\nu}
\def\t{\tau}
\def\cP{{\cal P}}

\def\l{\lambda}

\def\r{\rho}
\def\cL{{\cal L}}
\def\cA{{\cal A}}

\def\cA{{\cal A}}
\def\w{\omega}

\def\s{\sigma}
\def\e{\epsilon}

\def\bma{\begin{pmatrix}}
	\def\ema{\end{pmatrix}}

\def\bi{\begin{itemize}}
	\def\ei{\end{itemize}}

\title{\boldmath Physical content of Quadratic Gravity}

\preprint{IFT-UAM/CSIC-18-015 \ FTUAM-18-6}

\author[a]{Enrique Alvarez,}
\author[a]{Jesus Anero,}
\author[a]{Sergio Gonzalez-Martin} 
\author[a]{and Raquel Santos-Garcia}

\affiliation[a]{Departamento de F\'{\i}sica Te\'orica and Instituto de F\'{\i}sica Te\'orica, IFT-UAM/CSIC\\Universidad Aut\'onoma, 20849 Madrid, Spain}

\emailAdd{enrique.alvarez@uam.es} 
\emailAdd{jesusanero@gmail.com}
\emailAdd{sergio.gonzalezm@uam.es}
\emailAdd{raquel.santosg@uam.es}

\abstract{We have recently undergone an analysis of gravitational theories as defined in first order formalism, where the metric and the connection are treated as independent fields. The physical meaning of the connection field has historically been somewhat elusive. In this paper, a complete spin analysis of the torsionless connection field is performed, and its consequences are explored. The main properties of a hypothetical consistent truncation of the theory are discussed as well.
}
\begin{document}

	\maketitle
	\flushbottom
	\newpage
	\setcounter{page}{1}
	
	\section{Introduction}
	Theories of gravity where the lagrangian is quadratic in the Riemann tensor \cite{Stelle} are known to be well behaved in the ultraviolet (they are often asymptotically free) but suffer from  the fatal drawback of not being unitary (cf. \cite{Alvarez} for a general review, and \cite{AlvarezGaume} for a recent analysis similar in spirit to ours.)
	\par
	It has been recently pointed out \cite{AAG} that when considering those theories in first order formalism (which is not equivalent to the usual, second order one\footnote{Even for the  Einstein-Hilbert first order lagrangian the equivalence is lost as soon as fermionic matter is considered.}) where the metric and the connection are considered as independent physical fields, no quartic propagators appear and the theory is not obviously inconsistent. This framework is a good candidate for a unitary and renormalizable theory of the gravitational field, leading to a posible ultraviolet (UV) completion of General Relativity (GR). Recent work, following related lines, has been done regarding a possible UV completion of GR by modifying the usual second order quadratic gravity \cite{Einhorn:2017icw}\cite{Donoghue:2018izj}\cite{Salvio:2017qkx}. 
	\par
	Those theories depend on a number of independent coupling constants, which can be grouped into three big classes, corresponding to the Riemann tensor squared, the Ricci tensor squared, and the scalar curvature squared.
	
	There is one worrisome fact though. When considering the theory around a flat background there is no propagator for the graviton. This means that either the theory is not a theory of gravity at all, or else all the dynamics of the gravitational field is determined by the three index connection field. 
	\par
	Of course the idea that the true dynamics of gravitation is better  conveyed by the connection field than by the metric has a long history (cf. for example to the classic paper \cite{Ehlers}). It is the closest analogue to the usual gauge theories, and can be easily related to physical experiments and observations. In fact in \cite{AAG} we have shown that there are possible physical static connection sources that produce a $V(r)={C\over r}$ potential between them. This is at variance with what happens in the usual quadratic theories as formulated in second order, in which the natural potential is a scale invariant one $V(r)= C r$. This forces many authors  to include an Einstein-Hilbert (linear in the scalar curvature) piece in the action from the very beginning if one wants to reproduce solar-system observational constraints (cf \cite{Stelle} for a lucid discussion). Another possibility is a spotaneous symmetry breaking of the scale invariance of quadratic theories, so that the EH term is generated and dominates in the infrarred (see e.g \cite{Ferreira:2018itt}\cite{Einhorn:2015lzy}\cite{Salvio:2014soa}\cite{Shapiro:1994yt} regarding this issue). 
	\par
	The static connection  sources in \cite{AAG} were of the form $J_{\m\n\l}\sim j_\m T_{\n\l}+...$, where $j_\m$ was a conserved current and $T_{\m\n}$ was the energy-momentum tensor. The physical meaning of those sources is not clear, to say the least.
	In order to get a better grasp on the workings of the theory, it would be helpful to disentangle the  different physical spins contained  in the connection.	
	\par
	
	Our aim in this paper  is precisely  to perform a complete analysis of the physical content of the connection field. There are {\em a priori} 40 independent components in this field. We shall analyze them  by generalizing the usual spin projectors \cite{Barnes, VanNieuwenhuizen:1973fi} to the three-index case, and expanding the action in terms of  these projectors. We shall find that generically there is a spin 3 component, which disappears only when the coefficient of the Riemann square term vanishes. This property is however not stable with respect to quantum corrections, that will make this term reappear even if the classical coefficient is fine tuned to zero. Kinematically, there is also a set of three spin 2 components, five  spin 1 components and three spin 0 components.

	\par	
	Let us now summarize the contents of our paper. First we quickly review, mostly to establish our conventions, the spin content of the usual lagrangian linear in curvature (Einstein-Hilbert) both in second and in first order formalism. Then we tackle the spin analysis of theories quadratic in curvature, again both in second order and first order formalism. Extensive use is made of a new set of spin projectors, which are explained in the appendices.
	
	Throughout  this work we follow the Landau-Lifshitz spacelike conventions, in particular
	\be
	R^\m_{~\n\r\s}=\partial_\r \Gamma^\m_{\n\s}-\partial_\s \Gamma^\m_{\n\r}+\Gamma^\m_{\l\r}\Gamma^\l_{\n\s}-\Gamma^\m_{\l\s}\Gamma^\l_{\n\r}\ee
	and we define the Ricci tensor as
	\be R_{\m\n}\equiv R^\l_{~\m\l\n}\ee
	
	The commutator with our conventions is
	\bea[\nabla_\m,\nabla_\n]V^\l&=R^\l_{~\r\m\n}V^\r\nonumber\\
	[\nabla_\m,\nabla_\n]h^{\a\b}&=h^{\b\l} R^\a_{~\l\m\n}+h^{\a\l} R^\b_{~\l\m\n}\label{c}\eea

	\section{Lagrangians linear in curvature (Einstein-Hilbert) in second order formalism}
	Let us begin by quickly reviewing some well-known results on the quadratic (one loop) approximation of General Relativity (GR), as derived from the Einstein-Hilbert (EH) lagrangian. We do that mainly to establish our notation and methodology.
	
	We expand the EH action around flat space by taking
	\be
	g_{\m \n} = \eta_{\m \n} + \kappa h_{\m \n}
	\ee
	
	We are interested in the quadratic order of the expansion. The operator mediating the interaction between the metric perturbation reads
	\bea
	S&=\frac{1}{2}\int
	d^4 x~h^{\m\n}K^{\text{\tiny{EH}}}_{\m\n\r\s}h^{\r\s}\eea
where the operator reads
	\bea
	K^{\text{\tiny{EH}}}_{\m\n\r\s}\equiv &-{1\over 8}\left(\eta_{\m\r}\eta_{\n\s}+\eta_{\m\s}\eta_{\n\r}\right)\Box+ {1\over 8}\left(\pd_\m\pd_\r\eta_{\n\s}+\pd_\m\pd_\s \eta_{\n\r}-\pd_\n\pd_\r\eta_{\m\s}+\pd_\n\pd_\s \eta_{\m\r}\right)+\nonumber\\
	&-{1\over 4} \left(\pd_\r\pd_\s \eta_{\m\n}+\eta_{\r\s}\pd_\m\pd_\n\right)+ {1\over 4}\eta_{\m\n}\eta_{\r\s} \Box
	\label{Keh}
	\eea
	
	In order to better understand the physical content of this action, 
	we can decompose the symmetric tensor $h_{\m\n}$ as
	\be
	h_{\m\n}=h^2_{\m\n}+\Box^{-1}\left(\pd_\m A_\n+\pd_\n A_\m\right)-{\pd_\m\pd_\n\over \Box}\Phi+
	{1\over 3}\left(\eta_{\m\n}-{\pd_\m\pd_\n\over \Box}\right)\psi
	\ee
	where as we shall see $h^2_{\m\n}$ corresponds to the spin 2 part of the field. The other fields are defined as follows
	\bea
	&\phi\equiv \pd^\r\pd^\s h_{\r\s}\equiv \Box \Phi\nonumber\\
	&h\equiv \eta^{\m\n} h_{\m\n}\nonumber\\
	&A_\m\equiv \pd^\s h_{\m\s}; \quad \pd_\m A^\m= \Box \Phi
	\eea
 Under linearized diffeomorphisms 
 	\be
 	\d h_{\m\n}=\pd_\m\xi_\n+\pd_\n\xi_\m
 	\label{gauge}
 	\ee
 	 these  transform as
	\bea
	&\d \phi= 2 \Box^2 \xi\nonumber\\
	&\d h= 2 \Box \xi\nonumber\\
	&\d A_\m= \Box \xi_\m^T+ 2 \Box \pd_\m \xi
	\eea
	where we have split $\xi_{\m}$ in its transverse ($\xi_{\m}^T$) and longitudinal ($\partial_{\m} \xi$) parts.
	
	From the transformation properties, it is clear that there is a scalar gauge invariant combination
	\be
	\d \psi\equiv \d \left(h-\Phi\right)=0
	\ee
	
	As stated before, we want to carry out an analysis of the spin content of the fields in the theory using the spin projectors defined in Appendix  \ref{A}. The action of these spin projectors\footnote{It has to be understood that when writting the action of the projectors in terms of derivatives and box operators, it is implicit that these correspond to the ones of flat space.} over $h_{\m\n}$ gives
	\bea
	&h^{0 w}_{\m\n}\equiv (P_0^w h)_{\m\n}= \Box^{-2}\pd_\m\pd_\n \phi={\pd_\m\pd_\n \Phi\over \Box}; \quad\quad \d h^{0w}_{\m\n}= 2 \pd_\m\pd_\n\xi\nonumber\\
	&h^{0 s}_{\m\n}\equiv (P_0^s h)_{\m\n}={1\over 3}\bigg\{\eta_{\m\n}-{\pd_\m\pd_\n\over \Box}\bigg\}\psi;\quad\quad\d h^{0s}_{\m\n}=0\nonumber\\
	&h^1_{\m\n}\equiv (P_1 h)_{\m\n}=\Box^{-1}\left(\pd_\m A_\n+\pd_\n A_\m\right)-2 {\pd_\m\pd_\n \Phi\over \Box};\quad\quad\d h^1_{\m\n}=\pd_\m \xi_\n^T+\pd_\n \xi_\m^T\nonumber\\
	&h^2_{\m\n}\equiv (P_2 h)_{\m\n}=h_{\m\n}- \Box^{-1} \left(\pd_\m A_\n+\pd_\n A_\m\right)+\Box^{-2} \pd_\m\pd_\n \phi-\nonumber\\
	&\qquad-{1\over 3}\bigg\{h\eta_{\m\n}-\Box^{-1}\left(\pd_\m\pd_\n h+\phi \eta_{\m\n}\right)+\Box^{-2} \pd_\m\pd_\n \phi\bigg\}=\nonumber\\
	&\qquad=h_{\m\n}-\Box^{-1}\left(\pd_\m A_\n+\pd_\n A_\m\right)+{\pd_\m\pd_\n\over \Box}\Phi-
	{1\over 3}\left(\eta_{\m\n}-{\pd_\m\pd_\n\over \Box}\right)\psi; \quad\quad\d h^2_{\m\n}=0\label{P}
	\eea
	and integrating by parts we get
	\bea
	&\int  d(vol)\, h_{\m\n}^{0 s} \Box\, h^{\m\n}_{0 s}=\int d(vol)\, {1\over 3}\psi \Box \psi\nonumber\\
	&\int d(vol)\, (h_{\m\n}^{0s} + h_{\m\n}^{0w})  \Box\, (h^{\m\n}_{0s} + h^{\m\n}_{0w}) =\int d(vol)\, \left(\Phi \Box \Phi+{1\over 3} \psi \Box \psi\right)\nonumber\\
	&\int d(vol)\,h^1_{\m\n} \Box\, h_1^{\m\n}=\int d(vol)\,\left(-2 A_\m  A^\m -2 \Phi \Box \Phi\right)\nonumber\\
	&\int d(vol)\,h^2_{\m\n} \Box \,h_2^{\m\n}=\int d(vol)\,\left( h_{\m\n} \Box h^{\m\n}-{1\over 3}\psi \Box \psi+\Phi \Box \Phi+2 A_\m A^\m\right)
	\eea
	
	Then the Einstein-Hilbert action can be rewritten in terms of the projectors as
	\bea
	S^{\text{\tiny{EH}}}&=-\frac{1}{8}\int
	d^4 x~h^{\m\n}(P_2 - 2P_0^s)_{\m\n\r\s} \Box h^{\r\s} \label{EHp}
	\eea
	At this point, one can ask the question of whether it is possible to write a local lagrangian that contains only the spin 2 part of $h_{\m\n}$. Indeed the spin two part can be written as
	\bea
	&	h^2_{\m\n}=h_{\m\n}-{\partial_\m \partial^\r h_{\r\n}+h_{\m\r} \partial^\r \partial_\n\over \Box}+{\partial_\m \partial_\n \partial_\r \partial_\s h^{\r\s}\over \Box^2}-\nonumber\\
	&\qquad-{1\over 3}\bigg\{h\, \eta_{\m\n}-{\partial_\m \partial_\n\over \Box} h-\eta_{\m\n} {\partial^\r \partial^\s h_{\r\s}\over \Box}+ {\partial_\m \partial_\n \partial^\r \partial^\s h_{\r\s}\over \Box^2}\bigg\}
	\eea
	where we can see that we have a term which goes as $\tfrac{1}{\Box^2}$. This means that if we do not want to get non-local inverse powers of the d'Alembert operator, the simplest monomial that contains  spin 2 only is going to be given by
	\be
	S_2\equiv {1\over\kappa^6}\int d^4 x\, h^2_{\m\n} \Box^4 h_2^{\m\n}
	\label{h2}
	\ee
	which as is well-known suffers from several unitarity and causality problems associated to higher derivative lagrangians\footnote{Note that this action has a larger gauge symmetry, namely
		\be
		\d h_{\m\n}=\left(P_1\right)_{\m\n\r\s}\Lambda_1^{\r\s}+\left(P_0^s\right)_{\m\n\r\s}\Lambda_2^{\r\s}+\left(P_0^w\right)_{\m\n\r\s}\Lambda_3^{\r\s} \nonumber
		\ee
		where $\Lambda_i^{\m \n}$ are arbitrary fields.}.
	It would seem that the (harmless as we shall see) spin 0 addition is a necessary ingredient in a unitary Lorentz invariant spin 2 theory. We will come back to this point at the end of this work.
	\par
	Let us go back to the EH action \eqref{EHp}. With the help of \eqref{P}, we can further decompose it in terms of the different fields contained in $h_{\m\n}$
	\bea
	S^{\text{\tiny{EH}}}&=-\frac{1}{8}\int
	d^4 x~\left[h^{\m\n}\Box h_{\m\n}+2A_\m A^\m+\Phi\Box\Phi-\psi\Box\psi\right]\eea
	The equations of motion read
	\bea
	&\dfrac{\d S}{\d h^{\m\n}}=\Box h_{\m\n}=0\nonumber\\
	&{\d S\over \d \psi}=\Box\psi=0\nonumber\\
	&{\d S\over \d \Phi}= \Box \Phi=\phi=0 \nonumber\\
	&{\d S \over \d A_\m}= A_\m=0
	\eea
	so that $A_\m=\phi=0$, leaving just 5 free components in $h_{\m\n}$ on shell.

	In order to find  the propagator, we need to introduce a gauge fixing term to make \eqref{Keh} invertible. Let us choose the harmonic (de Donder) gauge condition given by the operator 	\bea
	K^{\text{\tiny{gf}}}_{\m \n \r \s} =& - \dfrac 1 8 \left(\partial_\m \partial_\r \eta_{\n \s}+\partial_\m \partial_\s \eta_{\n \r} + \partial_\n \partial_\r\eta_{\m \s} + \partial_\n \partial_\s \eta_{\m \r} \right)  -\dfrac 1 4\left(\eta_{\r\s}\pd_\m\pd_\n+ \eta_{\m \n} \partial _\r \partial _\s\right)-\nonumber\\
	& - \dfrac 1 8 \eta_{\m \n} \eta_{\r \s}\Box =- \dfrac{1}{4} \left(P_1 +\dfrac{3}{2} P_0^s + \dfrac 1 2 P_0^w - \dfrac{\sqrt{3}}{2} P^\times\right)_{\m\n\r\s} \ \Box
	\label{gfixing}
	\eea
	in such a way that
	\bea
	K^{\text{\tiny{EH+gf}}}_{\m \n \r \s} &= - \frac{1}{8} \left(\eta_{\m\r} \eta_{\n \s}+\eta_{\m\s}\eta_{\n\r}-\eta_{\m \n} \eta_{\r \s} \right) \ \Box =\nonumber\\
	&=- \dfrac{1}{4} \left(P_2 + P_1 - \dfrac{1}{2} P_0^s + \dfrac 1 2 P_0^w - \dfrac{\sqrt{3}}{2}P^\times\right)_{\m\n\r\s} \ \Box
	\eea
	
	The propagator is easily found to be
	\bea\label{propa}
	\Delta_{\m\n\r\s}&=- \frac{ 1}{ 4} \left(\eta_{\m\r} \eta_{\n \s}+\eta_{\m\s}\eta_{\n\r}-\eta_{\m \n} \eta_{\r \s} \right) \ \Box^{-1}=\nonumber\\
	&=-4\left(P_2+P_1-\frac{1}{2} P_0^s+\frac{1}{2}P_0^w-\frac{\sqrt{3}}{2}P^\times \right)_{\m\n\r\s} \Box^{-1}
	\eea
	
	We are also interested in computing the interaction energy between two external, conserved currents $T_{(1)}^{\m\n}$ and $T_{(2)}^{\m\n}$ 
	\be
	W\left[T_{(1)},T_{(2)}\right]=\int d^4 x T_{(1)}^{\m\n} \Delta_{\m\n\r\s} T_{(2)}^{\r\s}=\int d^4 x\left( T_{(1)}^{\m\n} \Box^{-1}  T_{(2)\m\n}-{1\over 2} T_{(1)} \Box^{-1} T_{(2)}\right)\label{EHW}
	\ee
	
	One may reasonably feel a little nervous about the negative sign of the spin 0 component in \eqref{EHp} as well as in \eqref{propa}. Let us demonstrate in a very explicit way that in spite of what it seems, the Einstein-Hilbert propagator is positive definite when saturated with physical sources.
	
	First we assume that massless gravitons are the carriers of the interaction. In momentum space we choose
	\be
	k^\m=\left(\kappa,0,0,\kappa\right)
	\ee
	and the conservation of energy-momentum implies
	\bea
	&T^{00}(k)= T^{33}(k)\nonumber\\
	&T^{0 i}(k)=T^{3 i}(k)
	\eea
	
	Then, an easy computation leads to the expression for the free energy in terms of the components of the two external conserved sources $T_{(1)}^{\m\n}$ and $T_{(2)}^{\m\n}$ as
	\bea
	&W\left[T_{(1)},T_{(2)}\right]=\int {d^4 k\over k^2}\bigg\{{1\over 2}\left(T_{(1)}^{11}-T_{(1)}^{22}\right)\left(T_{(2)}^{11}-T_{(2)}^{22}\right)+2 T_{(1)}^{12} T_{(2)}^{12}\bigg\}
	\eea
	which is positive semi-definite in case of identical sources $T_{(1)}^{\m\n}=T_{(2)}^{\m\n}$. 
	\par
	Moreover, for static sources the energy-momentum tensor reads (all other components vanish)
	\be
	T_{(1,2)}^{00}\equiv M_{(1,2)} \d^{(3)}\,\left(\vec{x}-\vec{x}_{(1,2)}\right)
	\ee
	and in momentum space
	\be
	T_{(1,2)}^{00}(k)\equiv M_{(1,2)}\, \d(k^0)\,e^{i\vec{k}\vec{x}_{(1,2)}}
	\ee
	it follows that
	\bea
	&W\left[T_{(1)},T_{(2)}\right]={1\over 2 C}\, M_1 M_2 \,\int {d^3 k\over \vec{k}^2}\,e^{i\vec{k}\left(\vec{x}_1-\vec{x}_2\right)}={\,\pi \over 2C}\,{M_1 M_2\over  \left|\vec{x}_1-\vec{x}_2\right|}
	\eea
	where we have represented
	\be
	\int d k_0\equiv {1\over C}
	\ee
	Therefore, the free energy is definite positive, as it should. 
	\section{Lagrangians linear in curvature in first order formalism}
	Let us now make the exercise of reanalyzing this same theory in first order formalism, in which the metric and the connection are independent. We shall find after some roundabout that the physical content of the theory is the same as we previously found in the last paragraph.
	
	We start with the Einstein-Hilbert action
	\be
	S^{\text{\tiny{EH}}}\equiv-{1\over 2 \kappa^2}~\int
	d^n x\sqrt{|g|}g^{\m\n}R_{\m\n}\left[\Gamma\right]
	\ee
	and we expand it around Minkowski spacetime as
	\bea
	&g_{\m\n}\equiv \eta_{\m\n}+\kappa h_{\m\n}\nonumber\\
	&\Gamma^\a_{\b\g}\equiv  A^\a_{\b\g}
	\eea
	where $A^\a_{\b\g}$ is the quantum field for the connection, which is symmetric in the last two indices as we are restricting ourselves to the torsionless case. 
	
	After this expansion the action can be written as
	\bea S^{\text{\tiny{EH}}}&=-\int
	d^n x\left\{h^{\g\e}N_{\g\e~\l}^{~~\a\b}A^{\lambda}_{\alpha\beta}+\frac{1}{2}A^{\t}_{\g\e}K^{\g\e~\a\b}_{~\t~~\l}A^{\lambda}_{\alpha\beta}\right\}\label{BF}
	\eea
	where the operators mediating the interactions have the form
	\bea
	N_{\g\e~\l}^{~~\a\b}&=\frac{1}{2\kappa}\left\{\frac{1}{2}\left(\eta_{\g\e}\eta^{\a\b}-\d^\a_\g\delta^{\beta}_{\e}-\d^\a_\e\delta^{\beta}_{\g}\right)\pd_\lambda
	-\right.\nonumber\\
	&\left.-\frac{1}{4}\left(\eta_{\g\e}\d^\b_\l\pd^\a+\eta_{\g\e}\d^\a_\l\pd^\b -\d^\a_\g\delta^{\beta}_{\lambda}\pd_\e-\d^\b_\g\delta^{\a}_{\lambda}\pd_\e-\d^\a_\e\delta^{\beta}_{\lambda}\pd_\g
	-\d^\b_\e\delta^{\a}_{\lambda}\pd_\g\right)\right\}  \\
	K^{\g\e~\a\b}_{~\t~~\l}&=\frac{1}{\kappa^2}\left\lbrace \frac{1}{4}[\d^\e_\t \d^\g_\l \eta^{\a\b}
	+\d^\g_\t \d^\e_\l \eta^{\a\b}+\d^\b_\l \d^\a_\t \eta^{\g\e} +\d^\a_\l \d^\b_\t \eta^{\g\e}
	\right.\nonumber\\ 
	&\left. -\d^\b_\t \d^\g_\l \eta^{\a\e}
	-\d^\b_\t \d^\e_\l \eta^{\a\g}
	-\d^\a_\t \d^\e_\l \eta^{\b\g}
	-\d^\a_\t \d^\g_\l \eta^{\b\e} ]
	\right\rbrace \nonumber \label{K}
	\eea
	
	From the path integral, the contribution to the effective action reads
	\bea
	e^{iW \scriptstyle\left[\eta_{\m\n}\right]}&=\int \mathcal{D}h\mathcal{D}A~e^{iS_{\text{\tiny FOEH}}[h,A]}\eea
	and using the background expansion (\ref{BF}) we can integrate over ${\cal D}A$ yielding
	\bea
	e^{iW}=\int \mathcal{D}h e^{\left\{-\frac{i}{2}\int
		d^n x~\sqrt{|g|}~\frac{1}{2}h^{\m\n}D_{\m\n\r\s}h^{\r\s}\right\}}\label{PI}\eea
	where
	\bea
	D_{\m\n\r\s}&=\frac{1}{4}(\eta_{\m\r}\eta_{\n\s}+\eta_{\m\s}\eta_{\n\r}-2\eta_{\m\n}\eta_{\r\s})\Box+\frac{1}{2}(\eta_{\m\n}\pd_\r \pd_\s+\eta_{\r\s}\pd_\m \pd_\n)\nonumber\\
	&-\frac{1}{8}(\eta_{\m\r}\pd_\n \pd_\s+\eta_{\m\s}\pd_\n \pd_\r+\eta_{\n\r}\pd_\m \pd_\s+\eta_{\n\s}\pd_\m \pd_\r)\nonumber\\
	&-\frac{1}{8}(\eta_{\m\r}\pd_\s \pd_\n+\eta_{\m\s}\pd_\r \pd_\n+\eta_{\n\r}\pd_\s \pd_\m+\eta_{\n\s}\pd_\r \pd_\m)
	\eea
	
	We now expand this operator in the basis of projectors (see Appendix \ref{A}) so that
	\bea
	D_{\m\n\r\s}&=\frac{1}{2}\left(P_2-(n-2)P_0^s\right)_{\m\n\r\s} \ \Box
	\eea
	and in this way the action can be rewritten (for $n=4$) as
	\bea
	S^{\text{\tiny{EH}}}&=-\frac{1}{8}\int
	d^4 x~h^{\m\n}(P_2 - 2P_0^s)_{\m\n\r\s} \Box h^{\r\s}\eea
	
	In conclusion, we obtain the same result when we treat the theory in second order formalism (\ref{EHp}) and  in first order formalism, for the particular case of the Einstein-Hilbert action.

	\section{Lagrangians quadratic in curvature  in second order formalism}
	
	Let us now begin the study of lagrangians quadratic in the spacetime curvature, first in the usual second order formalism.
	
	The most general action in this set (the connection is assumed  in this section to be the metric one) is
	\be
	S^{\text{\tiny{SOQ}}}\equiv~\int
	d^n x\sqrt{|g|}\left(\a R^2+\b R_{\m\n}R^{\m\n}+\g R_{\m\n\r\s}R^{\m\n\r\s}\right)
	\ee
	
	When we expand around flat space $g_{\m\n}=\eta_{\m\n}+\kappa h_{\m\n}$ it follows that 
	
	\bea\label{FG}
	S^\text{\tiny SOQ}=~\kappa^2\int
	d^n x h^{\m\n}&\Big\{\a\left[\pd_\m\pd_\n\pd_\r\pd_\s-\left(\eta_{\r\s}\pd_\m\pd_\n+\eta_{\m\n}\pd_\r\pd_\s\right)\Box+\eta_{\m\n}\eta_{\r\s}\Box^2\right]+\nonumber\\
	&+\frac{\b}{4}\left[2\pd_\m\pd_\n\pd_\r\pd_\s-\frac{1}{2}\left(\eta_{\m\r}\pd_\n\pd_\s+\eta_{\m\s}\pd_\n\pd_\r+\eta_{\n\r}\pd_\m\pd_\s+\eta_{\n\s}\pd_\m\pd_\r\right)\Box\right.\nonumber\\
	&\left.-\left(\eta_{\r\s}\pd_\m\pd_\n+\eta_{\m\n}\pd_\r\pd_\s\right)\Box+\frac{1}{2}\left(\eta_{\m\r}\eta_{\n\s}+\eta_{\m\s}\eta_{\n\r}\right)\Box^2+\eta_{\m\n}\eta_{\r\s}\Box^2\right]+\nonumber\\
	&+\frac{\g}{4}\left[4\pd_\m\pd_\n\pd_\r\pd_\s+2\left(\eta_{\m\r}\eta_{\n\s}+\eta_{\m\s}\eta_{\n\r}\right)\Box^2-\right.\nonumber\\
	&\left.-2\left(\eta_{\m\r}\pd_\n\pd_\s+\eta_{\m\s}\pd_\n\pd_\r+\eta_{\n\r}\pd_\m\pd_\s+\eta_{\n\s}\pd_\m\pd_\r\right)\Box\right]\Big\}h^{\r\s}
	\eea
	
	We can write the operator in terms of spin projectors as
	\bea K^{\text{\tiny{SOQ}}}_{\m\n\r\s}&=\kappa^2\left(\a(n-1)P_0^s+\frac{\b}{4}(P_2+nP_0^s)+\g(P_2+P_0^s)\right)_{\m\n\r\s}\ \Box^2=\nonumber\\
	&=\frac{\kappa^2}{4}\left(c_1P_2+c_2P_0^s\right)_{\m\n\r\s} \ \Box^2 \label{O}
	\eea
	where $c_1=\b+4\g$ and $c_2=4(n-1)\a+n\b+4\g$. 
	
	If we use the action of spin projectors over the graviton decomposition (\ref{P}), the action can be rewritten as
	\bea
	S^{\text{\tiny{SOQ}}}=&\frac{\kappa^2}{4}\int
	d^n x~ \left[c_1\left(h^{\m\n}\Box^2 h_{\m\n}+2A_\m\Box A^\m+\phi^2-\frac{1}{3}\psi\Box^2\psi\right)+\frac{c_2}{3}\psi\Box^2\psi\right]\label{SOQ}
	\eea
	
	Let us at this point make a short aside on the higher derivative scalar terms. Consider the lagrangian  \cite{Creminelli}
	\be
	L={1\over 2}\left(\pd_\m \psi\right)^2+ {1\over 2} C \psi\Box^2 \psi
	\ee
	and introduce an auxiliary field, $\chi$, so that
	\be
	L={1\over 2}\left(\pd_\m \psi\right)^2+  C\,\pd_\m\psi \pd^\m\chi-{1\over 2}C\, \chi^2
	\ee
	The EM for the auxiliary field just yields
	\be
	\chi=-\Box \psi
	\ee
	which just reproduces the original action. Now we can define
	\be
	\Psi\equiv \psi+C \chi
	\ee
	The mixing term disappears and the action diagonalizes to
	\be
	L={1\over 2}\left(\pd_\m  \Psi\right)^2- \dfrac 1 2 C^2 \left(\pd_\m\chi\right)^2 -{1\over 2}C\, \chi^2
	\ee
	
	It follows that the auxiliary field becomes a ghost no matter the value of the constant $C$. When there is no canonical kinetic term for the field $\psi$ this mechanism is not at work. However, such a term is always generated by the Einstein-Hilbert (linear in the space-time curvature) piece of the gravitational lagrangian. This linear piece is physically unavoidable, even if it is not present in the classical lagrangian, it will be generated by radiative corrections.\footnote{If we restrict ourselves only to the  $R^2$ terms, i.e. $\b=\g=0$, we get
		\bea
		S_\text{$R^2$}=&\kappa^2\a\int
		d^n x~ \psi\Box^2\psi\nonumber
		\eea
		so that the equation of motion reads
		\be\Box^2\psi=0\nonumber\ee
		From this we can see that there is a gauge invariant ghostly state.}
	
	Going back to our analysis, we can obtain the equations of motion for the quadratic action (\ref{SOQ})
	\bea
	&\dfrac{\d S}{\d h^{\m\n}}= c_1 \Box^2 h_{\m\n}=0\nonumber\\
	&\dfrac{\d S}{\d \psi}=(c_2-c_1)\Box^2 \psi=0\nonumber\\
	&\dfrac{\d S}{\d \phi}=c_1 \phi=c_1 \Box\Phi=0\nonumber\\
	&\dfrac{\d S}{\d A^{\m}}=c_1 \Box A_\m=0
	\eea
	
	Please note that the equations of motion have four derivatives so that the only way in which we can fix this problem is by taking $c_1=c_2=0$. This implies
	\bea
	&\b+4\g=\b+4\a=0
	\eea
	In this case the lagrangian is proportional to the Gauss-Bonnet density, i.e.  $\a=1,\b=-4,\g=1$ and $n=4$, and the operator (\ref{O}) reduces to
	\be K^{\text{\tiny{GB}}}_{\m\n\r\s}=0
	\ee
	This fact follows from the  identity
	\be 
	R^2 - 4R^{\mu\nu}R_{\mu\nu} + R^{\mu\nu\rho\sigma}R_{\mu\nu\rho\sigma}=\text{total derivative} 
	\ee

	
 Let us now  obtain the propagator for the general quadratic action \eqref{SOQ}, again in the harmonic gauge \eqref{gfixing} with a gauge parameter $-\tfrac{1}{2\xi}$. The operator reads
	\bea
	&K^{\text{\tiny{SOQ+gf}}}_{\m\n\r\s}=\frac{1}{8}\Bigg\{\frac{1}{\xi}P_1+2\kappa^2 c_1\Box P_2+\left(2\kappa^2c_2\Box+\frac{n-1}{2\xi}\right)P_0^s+\frac{1}{2\xi}P_0^w-\frac{\sqrt{n-1}}{2\xi}P_0^\times \Bigg\}_{\m\n\r\s}\Box
	\eea
	and inverting it we get
	\bea 
	&\Delta_{\m\n\r\s}\equiv (K^{-1})^{\text{\tiny{SOQ+gf}}}_{\m\n\r\s}=\frac{8}{k^2}\Bigg\{\xi P_1+\frac{1}{2\kappa^2c_1 k^2}P_2+\frac{\xi}{\kappa^2c_2 k^2}\left[\left(2\kappa^2c_2k^2+\frac{n-1}{2\xi}\right)P_0^w+\right.\nonumber\\
	&\left.+\frac{1}{2\xi}P_0^s+\frac{\sqrt{n-1}}{2\xi}P_0^\times \right]\Bigg\}_{\m\n\r\s}
	\eea
	provided $c_1\neq0$ and $c_2\neq0$.
	
	Now the interaction energy between external static sources, for $n=4$, is proportional to
	\be 
	W^{\text{\tiny{SOQ+gf}}}\propto T^{\m\n}\Delta^{\text{\tiny{SOQ+gf}}}_{\m\n\r\s}T^{\r\s}=\frac{4}{\kappa^2k^4}\left[\frac{1}{c_1}\left(T_{\m\n}T^{\m\n}-\frac{1}{3}T^2\right)+\frac{1}{3c_2}T^2\right]
	\ee
	
	This result is independent of the gauge fixing, and for the particular case $2c_1=-c_2$, the dependence on the sources is proportional to  the Einstein-Hilbert one 
	\be
	W^{\text{\tiny{SOQ+gf}}}\Big|_{c_2=-2c_1}\propto\frac{4}{\kappa^2k^4}\frac{1}{c_1}\left(T_{\m\n}T^{\m\n}-\frac{1}{2}T^2\right)
	\ee
	
	However, the factor $\frac{1}{k^4}$ in momentum space leads to a confining (linear) potential in position space.
	
	\subsection{Adding a term linear in the scalar curvature.}
	It has been argued in \cite{AAG} that a term linear in the spacetime curvature will be generated by quantum corrections, even if it is not initially present in the classical lagrangian. It is then of interest to consider the quadratic action plus the Einstein-Hilbert action
	\be
	S^{\text{\tiny{Q+EH}}}\equiv~\int
	d^n x\sqrt{|g|}\left(-{\l\over 2 \kappa^2}\,R+\a R^2+\b R_{\m\n}R^{\m\n}+\g R_{\m\n\r\s}R^{\m\n\r\s}\right)
	\ee 
	We can use the same harmonic gauge fixing \eqref{gfixing} with parameter $\xi$, so that the total operator can be written in terms of projectors as
	\bea K^{\text{\tiny{Q+EH+gf}}}_{\m\n\r\s}&=\frac{1}{8}\Bigg\{\frac{1}{\xi}P_1+(2\kappa^2c_1\Box+\l) P_2+\left(2\kappa^2c_2\Box+\frac{n-1}{2\xi}-\l(n-2)\right)P_0^s+\nonumber\\
	&+\frac{1}{2\xi}P_0^w-\frac{\sqrt{n-1}}{2\xi}P_0^\times\Bigg\}_{\m\n\r\s}\Box\eea
	Inverting the operator the propagator reads
	\bea {\Delta}^{\text{\tiny{Q+EH+gf}}}_{\m\n\r\s}&=\frac{8}{k^2}\Bigg\{\xi P_1+\frac{1}{2\kappa^2c_1 k^2+\l}P_2+\nonumber\\
	&+\frac{\xi}{\kappa^2c_2 k^2-\frac{\l(n-2)}{2}}\left[\left(2\kappa^2c_2k^2+\frac{n-1}{2\xi}-\l(n-2)\right)P_0^w+\frac{1}{2\xi}P_0^s+\frac{\sqrt{n-1}}{2\xi}P_0^\times \right]\Bigg\}_{\m\n\r\s}\eea\\
	
	Once we have the propagator, it is easy to check that the interaction energy between two external, static sources, for $n=4$, is proportional to
	\bea W&\propto T^{\m\n}(K^{-1})^{\text{\tiny{Q+EH+gf}}}_{\m\n\r\s}T^{\r\s}=\nonumber\\
	&=\frac{8}{\l}\left[\left(\frac{1}{k^2}-\frac{1}{(k^2+\frac{\l}{2\kappa^2c_1})}\right)\left(T_{\m\n}T^{\m\n}-\frac{1}{3}T^2\right)+\frac{2}{n-2}\left(\frac{1}{2(k^2-\frac{\l(n-2)}{2\kappa^2c_2})}-\frac{1}{2k^2}\right)\frac{T^2}{3}\right]=\nonumber\\
	&=\frac{8}{\l k^2}\left(T_{\m\n}T^{\m\n}-\frac{n-1}{3(n-2)}T^2\right)-\frac{8}{\l}\left[\frac{1}{(k^2+\frac{\l}{2\kappa^2c_1})}\left(T_{\m\n}T^{\m\n}-\frac{1}{3}T^2\right)-\frac{1}{2(k^2-\frac{\l(n-2)}{2\kappa^2c_2})}\frac{T^2}{3}\right]\eea
	
	Notice that the only contributions to the free energy come from $P_2$ and $P_0^s$ as the rest of spin operators do not contribute when saturated with the sources. The spin 2 piece can be rewritten as 
	\be \frac{8}{k^2(2\kappa^2c_1 k^2+\l)}P_2=\frac{8}{\l}\left[\frac{1}{k^2}-\frac{1}{(k^2+\frac{\l}{2\kappa^2c_1})}\right]P_2\ee\\
	The first term comes from the Einstein-Hilbert action, giving the well-known massless pole, whereas the second term corresponds to a massive $k^2=-\frac{\l}{2\kappa^2c_1}$ spin 2 pole with negative residue, coming from the quadratic action. 
	
	The spin 0 piece has the form
	\be \frac{8}{2k^2(\kappa^2c_2 k^2-\frac{\l(n-2)}{2})}P_0^s=\frac{16}{\l(n-2)}\left[\frac{1}{2(k^2-\frac{\l(n-2)}{2\kappa^2c_2})}-\frac{1}{2k^2}\right]P_0^s\ee\\
	
	In this case, the first term is a massive $k^2=\frac{\l(n-2)}{2\kappa^2c_2}$ spin 0 pole with positive residue, coming from the quadratic piece of the action. The second term is again the massless spin 0 pole with negative residue that we already encountered when studying the EH action.

	\section{Lagrangians quadratic in curvature in first order formalism}
Let us now enter into the main topic of this paper, namely the general situation in which the physics is conveyed by the graviton as well as by the connection field. Actually, as was pointed out in \cite{Alvarez}, when considering a metric fluctuating around flat space there is no kinetic term for the graviton, so that all the physics is encoded in the connection field. This is the main reason why we underwent a systematic analysis of the spin content of the said connection field. We consider the general action
\be
S_\text{\tiny FOQ}\equiv~\int
d^n x\sqrt{|g|}\left(\a R[\Gamma]^2+\b {R[\Gamma]}_{\m\n}{R[\Gamma]}^{\m\n}+\g  {R[\Gamma]}_{\m\n\r\s}{R[\Gamma]}^{\m\n\r\s}\right)
\ee
and we again use the expansion around Minkowski spacetime given by
\bea
&g_{\m\n}\equiv \eta_{\m\n}+\kappa h_{\m\n}\nonumber\\
&\Gamma^\a_{\b\g}\equiv  A^\a_{\b\g}
\eea
where $A^\a_{\b\g}$ is the quantum field for the connection, which is symmetric in the last two indices as we are restricting ourselves to the torsionless case. 

The action reduces to a kinetic term for the connection field
\bea S_{\text{\tiny FOQ}}&=\int
d^n xA^{\t}_{\m\n}K^{\m\n~\r\s}_{~\t~~\l}A^{\lambda}_{\r\s}
\eea
where the operator reads
\bea
K^{\m\n~\r\s}_{~\t~~\l}&=\a\Big\{\frac{1}{2}\left(\eta^{\m\n}\d^{\s}_{\t}\pd_\l\pd^{\r}+\eta^{\m\n}\d^{\r}_{\t}\pd_\l\pd^{\s}+\eta^{\r\s}\d^{\n}_{\l}\pd_\t\pd^{\m}+\eta^{\r\s}\d^{\m}_{\l}\pd_\t\pd^{\n}\right)-\eta^{\m\n}\eta^{\r\s}\pd_\l\pd_{\t}+\nonumber\\
&-\frac{1}{4}\left(
\d^{\n}_{\l}\d^{\s}_{\t}\pd^\m\pd^{\r}+\d^{\m}_{\l}\d^{\s}_{\t}\pd^\n\pd^{\r}+\d^{\n}_{\l}\d^{\r}_{\t}\pd^\m\pd^{\s}+\d^{\m}_{\l}\d^{\r}_{\t}\pd^\n\pd^{\s}\right)\Big\}+\nonumber\\
&+\b\Big\{\frac{1}{4}\left(\eta^{\m\r}\d^{\s}_{\t}\pd_\l\pd^{\n}+\eta^{\n\r}\d^{\s}_{\t}\pd_\l\pd^{\m}+\eta^{\m\s}\d^{\r}_{\t}\pd_\l\pd^{\n}+\eta^{\n\s}\d^{\r}_{\t}\pd_\l\pd^{\m}\right)+\nonumber\\
&+\frac{1}{4}\left(\eta^{\m\r}\d^{\n}_{\l}\pd_\t\pd^{\s}+\eta^{\n\r}\d^{\m}_{\l}\pd_\t\pd^{\s}+\eta^{\m\s}\d^{\n}_{\l}\pd_\t\pd^{\r}+\eta^{\n\s}\d^{\m}_{\l}\pd_\t\pd^{\r}\right)-\nonumber\\
&-\frac{1}{2}\left(\eta^{\m\r}\eta^{\n\s}+\eta^{\n\r}\eta^{\m\s}\right)\pd_\l\pd_{\t}-\frac{1}{4}\left(\eta^{\m\r}\d^{\n}_{\l}\d^{\s}_{\t}+\eta^{\n\r}\d^{\m}_{\l}\d^{\s}_{\t}+\eta^{\m\s}\d^{\n}_{\l}\d^{\r}_{\t}+\eta^{\n\s}\d^{\m}_{\l}\d^{\r}_{\t}\right)\Box\Big\}+\nonumber\\
&+\g\Big\{\eta_{\l\tau}\left[\frac{1}{2}\left(\eta^{\m\r}\pd^\s\pd^{\n}+\eta^{\n\r}\pd^\s\pd^{\m}+\eta^{\m\s}\pd^\r\pd^{\n}+\eta^{\n\s}\pd^\r\pd^{\m}\right)
-\left(\eta^{\m\r}\eta^{\n\s}+\eta^{\n\r}\eta^{\m\s}\right)\Box
\right]\Big\}
\eea

In the Appendix \ref{B} we have studied the spin projectors for connection fields $A\in \cA$, where $\cA$ is the space of torsionless connections (see Appendix \ref{C} for metric, torsionful connections). There are two main sectors in this space: the one corresponding to connections symmetric in the three indices (\ref{Bsym}), $A_S$, and the one endowed with the hook symmetry (\ref{Bhook}), $\cal{A}_H$, each one with 20 components. The spin content of the symmetric  sector is 
\be
\underline{20}_S=\left(\underline{3}\right)\oplus \left(\underline{2}\right)\oplus 2\,\left( \underline{1}\right)\oplus 2\,\left( \underline{0}\right)
\ee
and the spin content of the hook one is given by
\be
\underline{20}_H= 2 \left(\underline{2}\right)\oplus 3\,\left( \underline{1}\right)\oplus \,\left( \underline{0}\right)
\ee

There are 12 mutually orthogonal projectors on these different sectors. Projectors on the symmetric sector are represented by roman letters and indexed by the spin, $\text{P}_s$, whereas projectors in the hook sector are represented by calligraphic letters also indexed by the spin, ${\cal P}_s$. Nevertheless, this is not enough to expand the most general linear operator
\be
K: \cA\rightarrow \cA
\ee
which has dimension 22. In order to find a basis for this space, we need to add 10 new operators to the above set, which are not mutually orthogonal anymore. These new operators will be denoted as $\mathscr{P}_s$, where $s$ stands for the spin. Explicit expressions can be found in the Appendix \ref{Bmix}.
\par
Once we have obtained the complete basis for this space, we can expand the general operator in terms of these spin operators as

\bea
(K_{\text{FOQ}})^{\m\n~\r\s}_{~\t~~\l} &=\Big( -2 (2\g + \b) \ \text{P}_{0}^s -(4 \g + 9 \a + 2 \b) \ \mathcal{P}_0^{s} + (2 \g - \b)  \ \mathscr{P}_0^{x} - \dfrac{4}{3} (3 \g + 5 \b ) \ \text{P}_{1}^s \nonumber \\ 
&- 2 \g \ \mathcal{P}_1^{s}  - \dfrac{4}{3} (3 \g +  \b ) \ \mathcal{P}_1^{t} - (2 \g + \b) \ \mathscr{P}_1^{wx} + 4 \b \ \mathscr{P}_1^{ss} -2 (2 \g + \b ) \ (\text{P}_2 + \mathcal{P}_2)  \nonumber \\
&- 4 \g \ \mathcal{P}_{2}^{s} + 2 (\b + \g) \ \mathscr{P}_2^{x} - 4 \g \text{P}_3 \Big)^{\m\n~\r\s}_{~\t~~\l} \ \Box 
\label{KP} 
\eea

We also need to choose a gauge fixing, in this case we take
\be
S_{\text{gf}}= \dfrac{1}{\chi} \int
d^n x \ \eta^{\m \n} \eta^{\r\s} \eta_{\t\l} A^{\t}_{\m\n}A^{\lambda}_{\r\s} 
\ee
from where we can extract the operator which in terms of the projectors reads
\bea
(K_{\text{gf}})^{\m\n~\r\s}_{~\t~~\l} &= \dfrac{1}{\chi}\Big( \text{P}_0^w + 3 \ \text{P}_{0}^s + 3 \ \mathcal{P}_0^{s} - 3 \ \mathscr{P}_0^{x} + \mathscr{P}_0^{sw} + \mathscr{P}_0^{ws} + \text{P}_1 - \dfrac{5}{3} \ \text{P}_{1}^s + \mathcal{P}_{1}^{w} \nonumber \\
&+ \dfrac{2}{3} \ \mathcal{P}_1^{t} - \mathscr{P}_1^{wx} +\mathscr{P}_1^{ws} + \mathscr{P}_1^{sw} + \mathscr{P}_1^{sx} + 4 \ \mathscr{P}_1^{ss} \Big)^{\m\n~\r\s}_{~\t~~\l} \ \Box 
\eea

From the decomposition of the gauge fixing operator we see that the gauge fixing term does not posses any spin 2 or spin 3 piece. Looking at the operator \eqref{KP} for the three quadratic terms, we are going to have problems when $\gamma$ equals zero due to the fact that $\text{P}_3$, $\mathcal{P}_2^s$ and $\mathcal{P}_1^s$ dissapear from the scene. As we have seen, we cannot recover the spin 2 and spin 3 ones from the gauge fixing, so this leads to a non invertible operator, and thus, to new zero modes. 
\par
To understand this fact, let us  focus in the simplest case where $\b = \g = 0$. The operator for $R^2$ collapses to
\bea
(K_{\text{\tiny{$R^2$}}})^{\m\n~\r\s}_{~\t~~\l}&= - 9 \  (\mathcal{P}_0^{s})^{\m\n~\r\s}_{~\t~~\l} \ \Box\eea
so that

\vspace{-0.6cm}

\bea
(K_{ R^2+\text{gf}})^{\m\n~\r\s}_{~\t~~\l} &= \dfrac{1}{\chi}\Big( \text{P}_0^w + 3 \ \text{P}_{0}^s + (3 - 9 \chi ) \ \mathcal{P}_0^{s} - 3 \ \mathscr{P}_0^{x} + \mathscr{P}_0^{sw}+ \mathscr{P}_0^{ws} + \text{P}_1^w - \dfrac{5}{3} \ \text{P}_{1}^s \nonumber \\
&+ \mathcal{P}_{1}^{w} + \dfrac{2}{3} \ \mathcal{P}_1^{t} - \mathscr{P}_1^{wx}+\mathscr{P}_1^{ws} + \mathscr{P}_1^{sw} + \mathscr{P}_1^{sx} + 4 \ \mathscr{P}_1^{ss} \Big)^{\m\n~\r\s}_{~\t~~\l} \ \Box 
\eea

It follows that there are a grand total of 13 new zero modes. They are listed in the Appendix \ref{D}. Physically, this means that the theory has extra gauge symmetry when considered at one loop order, in addition to the one it has for the full theory, namely diffeomorphism and Weyl invariance. We are not aware of any other physical system where this happens. For what we can say, these extra gauge symmetries are accidental, and will disappear when computing higher loop orders. 

It is plain that the first order theory has a sector in which the connection reduces to the metric one. It is physically obvious that in this sector the theory should reduce to the one obtained in second order formalism.  Let us then check what happens when the connection reduces to the Levi-Civita connection. Around flat space we have
\be
A_{\m \n}^{\l~\text{(LC)}} = \partial_\m h_\n^\l +  \partial_\n h_\m^\l -  \partial^\l h_{\m\n}  
\ee

With this change we can extract an operator mediating interactions between the $h_{\m \n}$ and expand it in terms of the four-index spin projectors. In this way we can see how the six-index projectors and the four-index projectors talk to each other. The full correspondence is as follows

\newcolumntype{L}[1]{>{\raggedright\let\newline\\\arraybackslash\hspace{0pt}}m{#1}}
\newcolumntype{C}[1]{>{\centering\let\newline\\\arraybackslash\hspace{0pt}}m{#1}}
\newcolumntype{R}[1]{>{\raggedleft\let\newline\\\arraybackslash\hspace{0pt}}m{#1}}
\renewcommand{\arraystretch}{2}
\begin{tabular}{C{6cm} | C{6cm} }
	$\boldsymbol{ {A_{\l \m \n} P^{\l \m \n}_{\a \b \g} A^{\a \b \g}}}$ & $\boldsymbol{h_{\m \n} P ^{\m \n}_{\a \b} h^{\a \b}}$ \\  \hline 
	$\text{P}_0^w  $ & $ \dfrac{k^2}{4} P_0^w $ \\ \hline 
	$\text{P}_0^s $ & $ \dfrac{k^2}{36} (n-1) P_0^s$ \\  \hline 
	$ \mathcal{P}_0^s$ & $ \dfrac{2 k^2}{9} (n-1) P_0^s$ \\ \hline 
	$\mathscr{P}_0^x $ & $ \dfrac{ k^2}{6} (n-1) {P_0^s}$ \\  \hline 
	$\mathscr{P}_0^{sw}$ & $- \dfrac{ k^2}{3} \sqrt{n-1} P_0^\times$ \\ \hline 
	$\mathscr{P}_0^{ws}$ & $ \dfrac{ k^2}{12} \sqrt{n-1} P_0^\times $ \\  \hline 
	$\text{P}_1^{w} $ & $ \dfrac{ k^2}{6}  P_1$ \\ \hline 
	$\mathcal{P}_1^{w} $ & $\dfrac{ k^2}{3}  P_1$ \\  \hline 
	$\text{P}_2$ & $\dfrac{ k^2}{12}  P_2  -  \dfrac{ k^2}{36} (n-4)  P_0^s$ \\  \hline 
	$\mathcal{P}_2 $ & $ \dfrac{2 k^2}{3}  P_2-  \dfrac{ 2 k^2}{9} (n-4)  P_0^s$ \\ \hline 
	$\mathscr{P}_2^x $ & $ \dfrac{ k^2}{2}  P_2 - \dfrac{ k^2}{6} (n-4)  P_0^s$ \\  
\end{tabular}
\vspace{0.3cm}

where $\text{P}_1^s,\mathcal{P}_1^s,\mathcal{P}_1^w,\mathcal{P}_1^t,\mathscr{P}_1^{wx},\mathscr{P}_1^{ws},\mathscr{P}_1^{sw},\mathscr{P}_1^{sx},\mathscr{P}_1^{ss},\mathscr{P}_1^{wst},\mathcal{P}_2^s,\text{P}_3$, do not contribute when the connection reduces to the metric one.

The end result is that spin 3 collapses to zero, and the surviving different spin 2 sectors of the first order theory  degenerate into the unique spin 2 of the second order one.  Moreover, spin 1 reduces to spin 1 when going to second order formalism, as well as spin 0 goes to spin 0. 
\par
In the process however, a power of $k^2$ has been generated. This power is the responsible for the lack of (perturbative) unitarity of the theory in second order formalism. This problem then appears in this particular sector of the first order theory as well. 
\par
Then, unless a consistent method is found to isolate  this sector from the full first order theory ({\em id est}, a consistent truncation), the latter will inherit the unitarity problems of the second order one.	
	\section{Conclusions}
	When analyzing the connection field, one easily finds that there is generically a spin 3 component. This might be a problem in the sense that it is well-known (cf. for example \cite{Bekaert}) that it is not possible to build an interacting theory for spin 3 with a finite number of fields. Although we see no particular type of inconsistency to the order we have worked, it is always possible to avoid the presence of this spin 3 field altogether by choosing a particular set  of coupling constants, namely, putting to zero the coefficient of the Riemann squared term.
 This combination is not stable by renormalization, so that this choice implies a fine tuning of sorts. In addition there are several spin 0, spin 1 and spin 2 fields. This proliferation of spins occurs even for the Einstein-Hilbert action when in first order formalism.
 \par
 When the connection collapses to the metric (Levi-Civita) form, the spin 3 component disappears, and all spin 2 components are identified, but this sector suffers from the well-known unitarity problems present in second order formalism.
\par 
 In conclusion it is unclear whether it will be possible to define a truncation of the gravity lagrangian quadratic in curvature in first order formalism in which the problems of unitarity are absent. It seems that the healthy sectors do not describe gravity, and the sectors that do describe gravity fall into the known unitarity problems. To be specific, let us define a scalar product in $\cA$
 \be
 \langle A_1|A_2\rangle\equiv \int d(vol) A^1_{\m\n\l} A_2^{\m\n\l}
 \ee
 Then the subspace $\cA^\perp$ orthogonal to the metric connections
 \be
 A^{\text{(LC)}}_{\m\n\l}\equiv \pd_\m h_{\n\l}-\pd_\n h_{\l\m} -\pd_\l h_{\m\n}
 \ee
 is defined by
 \be
 A^\perp\in \cA^\perp\,\Leftrightarrow\, \partial^\m\left(A_{\m\n\l}-A_{\n\m\l}+A_{\l\n\m}\right)=0
 \ee
 which in terms of projectors reads
 \bea
 {A^{\perp}}_{\m\n\l} &=  {(\mathscr{P}_0^x)}_{\m\n\l}^{\r \s \t} \ \Omega_{\r \s \t}^1 +  {(\text{P}_1^s)}_{\m\n\l}^{\r \s \t} \  \Omega_{\r \s \t}^2 + {(\mathcal{P}_1^t)}_{\m\n\l}^{\r \s \t} \ \Omega_{\r \s \t}^3 +  {(\mathscr{P}_1^{ss})}_{\m\n\l}^{\r \s \t} \ \Omega_{\r \s \t}^4 \nonumber \\
 &+ {(\mathcal{P}_2^s)}_{\m\n\l}^{\r \s \t} \ \Omega_{\r \s \t}^5  + {(\mathscr{P}_2^x)}_{\m\n\l}^{\r \s \t} \ \Omega_{\r \s \t}^6+ {(\text{P}_3)} _{\m\n\l}^{\r \s \t}  \ \Omega_{\r \s \t}^7
  \eea
where $\Omega_{\r \s \t}^i \in {\cal A}$.
 
 Now, if we want to write a local lagrangian involving  $A^\perp$ only, we encounter the same problems we faced early on when we intended to write a lagrangian in terms of $h^2_{\m\n}$ only \eqref{h2}. For example, taking just the spin 3 part, due to the fact that ${(\text{P}_3)} _{\m\n\l}^{\r \s \t}  \ \Omega_{\r \s \t}$ goes as $\Box^{-3}$, we will need to have an action of the type
 \be
	S_3 ={1\over \kappa^{10}}\int d(vol) \, {A^{(3)}}_{\m\n\l} \Box^6 {A^{(3)}}^{\m\n\l}
 \ee
 if we want it to be  formally {\em local} (in the sense that no negative powers of $\Box$ appear).
 \par
 It is perhaps worth  remarking  that some of these problems are shared  even by  theories linear in curvature, as soon as fermionic matter is coupled to gravity. In this case the first order formalism and the second order one are not equivalent, and in fact when treating the theory in first order formalism, spacetime torsion is generated on shell. This fact seems worthy of some extra research.
 \par
 More work is clearly needed however before  a good understanding of the first order formalism is achieved. 

	\section*{Acknowledgments}
	One of us (EA) is grateful for discussions with prof. V. Berezin as well as for correspondence with M. Asorey, M. Einhorn, JM Gracia-Bondia and Joseph C. Varilly. We are grateful to G. Milans del Bosch for useful comments on the manuscript. RSG is supported by the Spanish FPU Grant No. FPU16/01595.
This work has received funding from the European Unions Horizon 2020 research and innovation programme under the Marie Sklodowska-Curie grants agreement No 674896 and No 690575. We also have been partially supported by FPA2016-78645-P(Spain), COST actions MP1405 (Quantum Structure of Spacetime) and  COST MP1210 (The string theory Universe). This work is supported by the Spanish Research Agency (Agencia Estatal de Investigacion) through the grant IFT Centro de Excelencia Severo Ochoa SEV-2016-0597.

	\newpage
	\appendix
	\section{Spin content and spin projectors}\label{A}

In order to get the spin projectors for a symmetric tensor $h_{\m\n}$, let us start with a simple vector field $u^\m$. If we consider a timelike reference momentum $k^\m$ (with $k^2>0$), physics is simpler in the adapted frame where 
\be
k^\m=\d^\m_0
\ee

Therefore, the spin content of a vector $u^\m$ which we represent as $\yng(1)$ is
\bea
\text{s=1}:\quad &u^i \quad \text{3 components},\nonumber\\
\text{s=0}:\quad & u^0\quad \text{1 component}.
\eea
And the corresponding projectors in momentum space read
\renewcommand{\arraystretch}{1}
\bea
&P^{(0)\b}_\a={k_\a k^\b\over k^2}\equiv \omega_\a\,^\b=\begin{pmatrix}1&0&0&0\\0&0&0&0\\0&0&0&0\\0&0&0&0\end{pmatrix}\nonumber\\
& P^{(1)\b}_\a= \d_\a^\b-{k_\a k^\b\over k^2}\equiv \theta_\a\,^\b=\begin{pmatrix}0&0&0&0\\0&1&0&0\\0&0&1&0\\0&0&0&1\end{pmatrix}
\eea
It should be noted that these operators are non-local in position space where
$\dfrac{1}{k^2}$ stands for $\Box^{-1}$. We shall use both momentum and position space as equivalent.
That is, we could as well write
\bea
&\omega_\a\,^\b={\pd_\a\pd^\b\over \Box}\nonumber\\
&\theta_\a\,^\b= \d_\a^\b-{\pd_\a\pd^\b\over \Box}
\eea
so the traces read as follows
\bea
&\text{Tr}~P_0=1\nonumber\\
&\text{Tr}~P_1=3 
\eea

As it is well-known, the metric $h_{\m\n}$ (or equivalently, the frame field, $h^a\,_\m$) transforms in the euclidean setting under the representation $\underline{10}\equiv \yng(2)$ of SO(4), so the spin content and corresponding projectors are given by
\bea
\text{s=2}:\quad & h^T_{ij}\equiv h_{ij}-{1\over 3} h\d_{ij}\quad &\quad &\left(P_2\right)_{\m\n}^{\r\s}\equiv {1\over 2}\left(\theta_\m^\r\theta_\n^\s+\theta_\m^\s\theta_\n^\r\right)-{1\over 3}\theta_{\m\n}\theta^{\r\s}\nonumber\\
\text{s=1}:\quad & h_{0i}&\nonumber \quad &\left(P_1\right)_{\m\n}^{\r\s}\equiv{1\over 2}\left(\theta_\m^\r\omega_\n^\s+\theta_\m^\s\omega_\n^\r+\theta_\n^\r\omega_\m^\s+\theta_\n^\s\omega_\m^\r\right)\\
\text{s=0}:\quad & h_{00}&\quad &\left(P_0^w\right)_{\m\n}^{\r\s}\equiv \omega_{\m\n}\omega^{\r\s}\nonumber\\
\text{s=0}:\quad & h\equiv \d^{ij} h_{ij}&\quad &\left(P_0^s\right)_{\m\n}^{\r\s}\equiv {1\over 3}\theta_{\m\n}\theta^{\r\s}
\eea
These particular projectors have been studied previously by Barnes and Rivers \cite{Barnes}. They are complete in the symmetrized direct product
\be
Sym\left(T_x\otimes T_x\right)
\ee
where $T_x$ is the tangent space at the point $x\in M$ of the space-time manifold.
\par
It is convenient to define another projector
\be
P_0\equiv P_0^w+P_0^s
\ee
and the non-differential projectors are
\bea
&I_{\m\n}^{\r\s}\equiv{1\over 2}\left(\d_\m^\r\d_\n^\s+\d_\m^\s \d_\n^\r\right)\nonumber\\
&T_{\m\n}^{\r\s}\equiv {1\over 4} \eta_{\m\n} \eta^{\r\s}
\eea
Then we can write a closure relation for these projectors, to be specific,
\be
\left(P_2\right)_{\m\n}^{\r\s}+\left(P_1\right)_{\m\n}^{\r\s}+\left(P_0\right)_{\m\n}^{\r\s}=I_{\m\n}^{\r\s}
\ee
These projectors are not enough though, as they do not form a base of the space of four-index tensors of the type of interest. Such a base is formed by five independent monomials, namely (permutations are implicit)
\bea
&M_1\equiv k_\m k_\n k_\r k_\s\nonumber\\
& M_2\equiv  k_\m k_\n \eta_{\r\s}\nonumber\\
&M_3\equiv k_\m k_\s \eta_{\r\n}\nonumber\\
&M_4\equiv \eta_{\m\n}\eta_{\r\s}\nonumber\\
&M_5\equiv \eta_{\m\r}\eta_{\n\s}
\eea
Therefore, in order to get a basis, we then need to add a new independent operator
\be
\left(P_0^\times\right)_{\m\n}^{\r\s}={1\over \sqrt{3}}\left(\omega_{\m\n}\theta^{\r\s}+\theta_{\m\n} \omega^{\r\s}\right)\label{PX}
\ee

that can be identified with the mixing of the two spin 0 components, $h$ and $h_{00}$.
It is clear that this new operator cannot be orthogonal to the other four, since closure implies that the only operator orthogonal to the set that closes is the null operator.

\section{Spin content of the  symmetric connection field}\label{B}

In this appendix, we decompose the operators mediating between two connection fields $A_{\m\b\g}\equiv g_{\a\m}\Gamma^\a_{\b\g}$ --symmetric in the last two indices, because we are assuming vanishing torsion-- in terms of the spin projectors of this field. The procedure is analogue to the one followed in Appendix \ref{A}.

Since  $A_{\m\n\l}=A_{\m\l\n}$,
\be
A_{\m\n\l}\in {\cal A}\equiv T_x\otimes Sym\left(T_x\otimes T_x\right)
\ee
The quadratic kinetic operator in this space is
\be
K\in {\cal A}\otimes{\cal A}
\ee
In order to disentangle the physical meaning of the gauge piece of the total action, we would like to expand $K$ as a sum of projectors with definite spin. There are 22 independent monomials to consider. Let un proceed by steps.
\par
The projector into ${\cal A}$ --namely, the identity in this space-- is
\bea
&P_0\equiv (P_0)_{\m(\n\l)}^{\a(\b\g)}\equiv {1\over 2}\d_\m^\a\left(\d_\n^\b \d_\l^\g+\d_\n^\g \d_\l^\b\right)={1\over 2}\left(1,0,0,1,0,0\right)\nonumber\\
&P_0^2\equiv (P_0)_{\m(\n\l)}^{\a(\b\g)}(P_0)_{\a(\b\g)}^{a(bc)}=P_{\m(\n\l)}^{a(bc)}=\cP_0\nonumber\\
&P_0 \,{\cal A}={\cal A}
\eea
(where the last equality in the first equation  refers to the vector notation introduced in the Appendix \ref{E}).
The subspace ${\cal A}$ corresponds, in terms of representations of the tangent group SO(4), to the sum of a totally symmetric three-index tensor plus a tensor with the {\em hook} symmetry
\be
\{2,0\}\otimes \{1\}=\{3,0\}\oplus \{2,1\} \quad\quad \yng(2)\otimes \yng(1)= \yng(3)\oplus \yng(2,1)
\ee

In terms of dimensions this is $\underline{40}=\underline{20}+\underline{20}$. The Young projectors are
\bea
&\text{ P}_S\equiv \left(\text{P}_{\tiny{\young(\a\b\g)}}\right)_{\m\n\l}^{\a\b\g}\equiv {1\over 6}\bigg\{\d^\a_\m \d^\b_\n \d^\g_\l+\d^\b_\m\d^\g_\n\d^\a_\l+ \d^\g_\m\d^\a_\n\d^\b_\l+ \d^\a_\m\d^\g_\n\d^\b_\l+ \d^\b_\m\d^\a_\n\d^\g_\l+\d^\g_\m\d^\b_\n\d^\a_\l\bigg\}=\nonumber\\
&={1\over 6}\left(1,1,1,1,1,1\right)
\eea
and the hook representation
\bea
\cP_H\equiv \left(P_{\tiny{\young(\a \b,\g)}}\right)_{\m\n\l}^{\a\b\g}&\equiv {1\over 3}\bigg\{\d^\a_\m\d^\b_\n \d^\g_\l+\d^\a_\m\d_\n^\g\d^\b_\l -\dfrac{1}{2}\d^\a_\n\d^\b_\m\d^\g_\l-\dfrac{1}{2}\d^\a_\n\d^\b_\l\d^\g_\m-\dfrac{1}{2}\d^\a_\l\d^\b_\n\d^\g_\m-\dfrac{1}{2}\d^\a_\l\d^\b_\m\d^\g_\n\bigg\}=\nonumber\\
&={1\over 3}\left(1,-\dfrac{1}{2},-\dfrac{1}{2},1,-\dfrac{1}{2},-\dfrac{1}{2}\right)
\eea

It should be stressed that this projector is not symmetric in $(\a\b)$, but rather in $(\b,\g)$. 

\bea
&\left(\cP_{\tiny{\young(\a \b,\g)}}\right)_{\m\n\l}^{\a\b\g}=\left(\cP_{\tiny{\young(\a \g,\b)}}\right)_{\m\n\l}^{\a\b\g}\nonumber\\
&\left(\cP\,_{\tiny{\young(\a \b,\g)}}\right)_{\m\n\l}^{\a\b\g}+\left(\cP\,_{\tiny{\young(\g \a,\b)}}\right)_{\m\n\l}^{\a\b\g}+\left(\cP\,_{\tiny{\young(\b \g,\a)}}\right)_{\m\n\l}^{\a\b\g}=0
\eea

In the following, we will keep this notation: $\text{P}$ for the projectors in the symmetric subspace and $\cP$ for those in the hook subspace.\newline

The Young projectors are symmetric, orthogonal and add to the identity in ${\cal A}$
\bea
&\text{P}^T_{S}=\text{P}_{S}\qquad\cP^T_{H}=\cP_{H}\nonumber\\
&\text{P}_S \cP_H=\cP_H \text{P}_S=0\nonumber\\
&{\text P}_S+{\cal P}_H= P_0
\eea
Then we can always write for any $A\in \cA$
\be
A=\cP_0 A=\text{A}_S+{\cal A}_H
\ee
with
\bea
&\text{P}_S \text{A}_S=\text{A}_S\nonumber\\
&\mathcal{P}_H {\cal A}_H={\cal A}_H
\eea

\subsection{The totally symmetric tensor}\label{Bsym}
Let us start by determining  the spin content of the totally symmetric piece $\left(P_{\{3\}} A \right)_{\a\b\g}\equiv A_{(\a\b\g)}$. 

We can decompose it in its spin components as
\bi
\item First the spin 3 component, which is given in the rest frame by
\be
\quad  A^T_{ijk}\equiv A_{ijk}-{1\over 5}\left(A_i\d_{jk}+A_j\d_{ik}+A_k\d_{ij}\right)
\ee
where
\be
A_i\equiv \sum_j A_{i jj}
\ee
There are of course 7 components in this set.

The spin 3 projector reads
\bea
(\text{P}_3)_{\l \m \n}^{\a \b \g}&=\dfrac{1}{6}\bigg( \theta^{\a}{}_{\n} \theta^{\b}{}_{\m} \theta^{\g}{}_{\l} + \theta^{\a}{}_{\m} \theta^{\b}{}_{\n} \theta^{\g}{}_{\l} +  \theta^{\a}{}_{\n} \theta^{\b}{}_{\l} \theta^{\g}{}_{\m} +  \theta^{\a}{}_{\l} \theta^{\b}{}_{\n} \theta^{\g}{}_{\m} + \theta^{\a}{}_{\m} \theta^{\b}{}_{\l} \theta^{\g}{}_{\n} +  \theta^{\a}{}_{\l} \theta^{\b}{}_{\m} \theta^{\g}{}_{\n}\bigg)\nonumber\\
& -  \dfrac{1}{15}\bigg( \theta^{\a}{}_{\n} \theta^{\b\g} \theta_{\m\l}+ \theta^{\a\g} \theta^{\b}{}_{\n} \theta_{\m\l}+\theta^{\a\b} \theta^{\g}{}_{\n} \theta_{\m\l}+\theta^{\a}{}_{\m} \theta^{\b\g} \theta_{\n\l}+ \theta^{\a\g} \theta^{\b}{}_{\m} \theta_{\n\l} +\theta^{\a\b} \theta^{\g}{}_{\m} \theta_{\n\l} +\nonumber\\
& +  \theta^{\a}{}_{\l} \theta^{\b\g} \theta_{\m\n}+\theta^{\a\g} \theta^{\b}{}_{\l} \theta_{\m\n} +\theta^{\a\b} \theta^{\g}{}_{\l} \theta_{\m\n}\bigg)
\eea
\item The  spin 2 component is given in the rest frame by
\be
A^T_{0ij}\equiv A_{0ij}-{1\over 3} A_0\d_{ij}
\ee
where
\be
A_0\equiv \sum_i  A_{0 ii}
\ee
The projector reads
\bea
(\text{P}_2)_{\l \m \n}^{\a \b \g}=&\tfrac{1}{6} \theta^{\beta}{}_{\nu} \theta^{\gamma}{}_{\mu} \omega^{\alpha}{}_{\lambda} + \tfrac{1}{6} \theta^{\beta}{}_{\mu} \theta^{\gamma}{}_{\nu} \omega^{\alpha}{}_{\lambda} -  \tfrac{1}{9} \theta^{\beta \gamma} \theta_{\mu \nu} \omega^{\alpha}{}_{\lambda} + \tfrac{1}{6} \theta^{\beta}{}_{\nu} \theta^{\gamma}{}_{\lambda} \omega^{\alpha}{}_{\mu} + \tfrac{1}{6} \theta^{\beta}{}_{\lambda} \theta^{\gamma}{}_{\nu} \omega^{\alpha}{}_{\mu} \nonumber \\ 
& -  \tfrac{1}{9} \theta^{\beta \gamma} \theta_{\lambda \nu} \omega^{\alpha}{}_{\mu} + \tfrac{1}{6} \theta^{\beta}{}_{\mu} \theta^{\gamma}{}_{\lambda} \omega^{\alpha}{}_{\nu} + \tfrac{1}{6} \theta^{\beta}{}_{\lambda} \theta^{\gamma}{}_{\mu} \omega^{\alpha}{}_{\nu} -  \tfrac{1}{9} \theta^{\beta \gamma} \theta_{\lambda \mu} \omega^{\alpha}{}_{\nu} + \tfrac{1}{6} \theta^{\alpha}{}_{\nu} \theta^{\gamma}{}_{\mu} \omega^{\beta}{}_{\lambda} \nonumber \\ 
& + \tfrac{1}{6} \theta^{\alpha}{}_{\mu} \theta^{\gamma}{}_{\nu} \omega^{\beta}{}_{\lambda} -  \tfrac{1}{9} \theta^{\alpha \gamma} \theta_{\mu \nu} \omega^{\beta}{}_{\lambda} + \tfrac{1}{6} \theta^{\alpha}{}_{\nu} \theta^{\gamma}{}_{\lambda} \omega^{\beta}{}_{\mu} + \tfrac{1}{6} \theta^{\alpha}{}_{\lambda} \theta^{\gamma}{}_{\nu} \omega^{\beta}{}_{\mu} -  \tfrac{1}{9} \theta^{\alpha \gamma} \theta_{\lambda \nu} \omega^{\beta}{}_{\mu} \nonumber \\ 
& + \tfrac{1}{6} \theta^{\alpha}{}_{\mu} \theta^{\gamma}{}_{\lambda} \omega^{\beta}{}_{\nu} + \tfrac{1}{6} \theta^{\alpha}{}_{\lambda} \theta^{\gamma}{}_{\mu} \omega^{\beta}{}_{\nu} -  \tfrac{1}{9} \theta^{\alpha \gamma} \theta_{\lambda \mu} \omega^{\beta}{}_{\nu} + \tfrac{1}{6} \theta^{\alpha}{}_{\nu} \theta^{\beta}{}_{\mu} \omega^{\gamma}{}_{\lambda} + \tfrac{1}{6} \theta^{\alpha}{}_{\mu} \theta^{\beta}{}_{\nu} \omega^{\gamma}{}_{\lambda} \nonumber \\ 
& -  \tfrac{1}{9} \theta^{\alpha \beta} \theta_{\mu \nu} \omega^{\gamma}{}_{\lambda} + \tfrac{1}{6} \theta^{\alpha}{}_{\nu} \theta^{\beta}{}_{\lambda} \omega^{\gamma}{}_{\mu} + \tfrac{1}{6} \theta^{\alpha}{}_{\lambda} \theta^{\beta}{}_{\nu} \omega^{\gamma}{}_{\mu} -  \tfrac{1}{9} \theta^{\alpha \beta} \theta_{\lambda \nu} \omega^{\gamma}{}_{\mu} + \tfrac{1}{6} \theta^{\alpha}{}_{\mu} \theta^{\beta}{}_{\lambda} \omega^{\gamma}{}_{\nu} \nonumber \\ 
& + \tfrac{1}{6} \theta^{\alpha}{}_{\lambda} \theta^{\beta}{}_{\mu} \omega^{\gamma}{}_{\nu} -  \tfrac{1}{9} \theta^{\alpha \beta} \theta_{\lambda \mu} \omega^{\gamma}{}_{\nu}\eea

\item There are two spin 1 components. First the one that is given in the rest frame by
\be
A_{ijk}\d^{jk}
\ee
with projector
\bea
(\text{P}_{1}^s)_{\l \m \n}^{\a \b \g}&=\dfrac{1}{15}\bigg( \theta^{\a}{}_{\n} \theta^{\b\g} \theta_{\m\l} + \theta^{\a\g} \theta^{\b}{}_{\n} \theta_{\m\l} +  \theta^{\a\b} \theta^{\g}{}_{\n} \theta_{\m\l} +  \theta^{\a}{}_{\m} \theta^{\b\g} \theta_{\l\n} + \theta^{\a\g} \theta^{\b}{}_{\m} \theta_{\l\n} +  \theta^{\a\b} \theta^{\g}{}_{\m} \theta_{\l\n} +\nonumber\\
&+  \theta^{\a}{}_{\l} \theta^{\b\g} \theta_{\m\n} + \theta^{\a\g} \theta^{\b}{}_{\l} \theta_{\m\n} +  \theta^{\a\b} \theta^{\g}{}_{\l} \theta_{\m\n}\bigg)
\eea
The other corresponds to
\be
A_{00i}
\ee
and the projector is
\bea
(\text{P}_1^w)_{\l \m \n}^{\a \b \g}&=\dfrac{1}{6}\bigg( \theta^{\g}{}_{\n} w^{\a}{}_{\m} w^{\b}{}_{\l} +  \theta^{\g}{}_{\m} w^{\a}{}_{\n} w^{\b}{}_{\l} +  \theta^{\g}{}_{\n} w^{\a}{}_{\l} w^{\b}{}_{\m} + \theta^{\g}{}_{\l} w^{\a}{}_{\n} w^{\b}{}_{\m} +  \theta^{\g}{}_{\m} w^{\a}{}_{\l} w^{\b}{}_{\n} \nonumber\\
&+  \theta^{\g}{}_{\l} w^{\a}{}_{\m} w^{\b}{}_{\n} + \theta^{\b}{}_{\n} w^{\a}{}_{\m} w^{\g}{}_{\l} +  \theta^{\b}{}_{\m} w^{\a}{}_{\n} w^{\g}{}_{\l} +  \theta^{\a}{}_{\n} w^{\b}{}_{\m} w^{\g}{}_{\l} + \dfrac{1}{6} \theta^{\a}{}_{\m} w^{\b}{}_{\n} w^{\g}{}_{\l} \nonumber\\
&+  \theta^{\b}{}_{\n} w^{\a}{}_{\l} w^{\g}{}_{\m} +  \theta^{\b}{}_{\l} w^{\a}{}_{\n} w^{\g}{}_{\m} + \dfrac{1}{6} \theta^{\a}{}_{\n} w^{\b}{}_{\l} w^{\g}{}_{\m} +  \theta^{\a}{}_{\l} w^{\b}{}_{\n} w^{\g}{}_{\m} +  \theta^{\b}{}_{\m} w^{\a}{}_{\l} w^{\g}{}_{\n} \nonumber\\
&+  \theta^{\b}{}_{\l} w^{\a}{}_{\m} w^{\g}{}_{\n} + \dfrac{1}{6} \theta^{\a}{}_{\m} w^{\b}{}_{\l} w^{\g}{}_{\n} +  \theta^{\a}{}_{\l} w^{\b}{}_{\m} w^{\g}{}_{\n}\bigg)
\eea
\item There are also two different  spin zero components. The first one corresponds to 
\be
A_{000}
\ee
and its projector is
\bea
(\text{P}_0^w)_{\l \m \n}^{\a \b \g}&=\dfrac{1}{6}\bigg( \omega^{\a}{}_{\n} \omega^{\b}{}_{\m} \omega^{\g}{}_{\l} + \omega^{\a}{}_{\m} \omega^{\b}{}_{\n} \omega^{\g}{}_{\l} +  \omega^{\a}{}_{\n} \omega^{\b}{}_{\l} \omega^{\g}{}_{\m} +  \omega^{\a}{}_{\l} \omega^{\b}{}_{\n} \omega^{\g}{}_{\m} +\nonumber\\
&+  \omega^{\a}{}_{\m} \omega^{\b}{}_{\l} \omega^{\g}{}_{\n} +  \omega^{\a}{}_{\l} \omega^{\b}{}_{\m} \omega^{\g}{}_{\n}\bigg)
\eea
while the second one corresponds to 
\be
A_{0ij}\d^{ij}
\ee
with projector
\bea
(\text{P}_{0}^s)_{\l \m \n}^{\a \b \g}&=\dfrac{1}{9}\bigg( \theta^{\b\g} \theta_{\m\n} w^{\a}{}_{\l} +  \theta^{\b\g} \theta_{ln} w^{\a}{}_{\m} +  \theta^{\b\g} \theta_{\m\l} w^{\a}{}_{\n} +  \theta^{\a\g} \theta_{\m\n} w^{\b}{}_{\l} +  \theta^{\a\g} \theta_{\l\n} w^{\b}{}_{\m} \nonumber\\
&+  \theta^{\a\g} \theta_{\m\l} w^{\b}{}_{\n} +  \theta^{\a\b} \theta_{\m\n} w^{\g}{}_{\l} +  \theta^{\a\b} \theta_{\l\n} w^{\g}{}_{\m} +  \theta^{\a\b} \theta_{\m\l} w^{\g}{}_{\n}\bigg)
\eea 
\par
Altogether we have accounted for the 20 components in this set and the spin content is
\be
\underline{20}_S=\left(\underline{3}\right)\oplus \left(\underline{2}\right)\oplus 2\,\left( \underline{1}\right)\oplus 2\,\left( \underline{0}\right)
\ee

Indeed, they satisfy the closure relation that symbolicall reads, 

\be\text{P}_0^s+\text{P}_0^w+\text{P}_1^s+\text{P}_1^w+\text{P}_2+\text{P}_3=\text{P}_S\label{closureS}\ee
\ei
\subsection{The hook sector}\label{Bhook}
Let us now work out the spin content of the 20 components of the diagram  $P_{\{2,1\}} A$.
\par
We will henceforth assume that connections are already projected into the corrresponding Young subspace, that is, when $A\in \cA$,
\be
{\cal A}^H_{\a\b\g}\equiv \left(\cP_H A\right)_{\a\b\g}\equiv {1\over 3}\left(2 A_{\a\b\g}- A_{\b\g\a}- A_{\g\a\b}\right)=A_{\a\b\g}
\ee
This implies cyclic symmetry
\be
{\cal A}_{\a\b\g}+{\cal A}_{\b\g\a}+{\cal A}_{\g\a\b}=0
\ee

Consider first components with one element in the direction of the momentum (that is the 0-th component in the rest frame). Remember that for the projectors acting in this subspace we are using the letter $\mathcal{P}$.
\bi
\item
There is only one spin zero, a trace that is given by
\be 
\sum_{i=1}^3 A_{i 0 i}
\ee
that is
\bea
(\mathcal{P}_0^{s})_{\l \m \n}^{\a \b \g}&=- \dfrac{1}{9} \theta^{\b\g} \theta_{\m\n} w^{\a}{}_{\l} + \dfrac{2}{9} \theta^{\b\g} \theta_{\n\l} w^{\a}{}_{\m} -  \dfrac{1}{9} \theta^{\b\g} \theta_{\m\l} w^{\a}{}_{\n} + \dfrac{1}{18} \theta^{\a\g} \theta_{\m\n} w^{\b}{}_{\l} \nonumber\\
&-  \dfrac{1}{9} \theta^{\a\g} \theta_{\n\l} w^{\b}{}_{\m} + \dfrac{1}{18} \theta^{\a\g} \theta_{\m\l} w^{\b}{}_{\n} + \dfrac{1}{18} \theta^{\a\b} \theta_{\m\n} w_{\l}{}^{\g} \nonumber\\
&-  \dfrac{1}{9} \theta^{\a\b} \theta_{\n\l} w_{\m}{}^{\g} + \dfrac{1}{18} \theta^{\a\b} \theta_{\m\l} w_{\n}{}^{\g}
\eea
\item There are three  spin 1 components. First 
\be
{1\over 2}\left(A_{j 0 i}-A_{i 0 j}\right)
\ee  
corresponding to
\bea
(\mathcal{P}_1^{s})_{\l \m \n}^{\a \b \g}&=- \dfrac{1}{4} \theta^{\a}{}_{\n} \theta_{\m}{}^{\g} w^{\b}{}_{\l} + \dfrac{1}{4} \theta^{\a}{}_{\m} \theta_{\n}{}^{\g} w^{\b}{}_{\l} + \dfrac{1}{4} \theta^{\a}{}_{\m} \theta_{\l}{}^{\g} w^{\b}{}_{\n} -  \dfrac{1}{4} \theta^{\a}{}_{\l} \theta_{\m}{}^{\g} w^{\b}{}_{\n} -  \dfrac{1}{4} \theta^{\a}{}_{\n} \theta^{\b}{}_{\m} w_{\l}{}^{\g} +\nonumber\\
&+ \dfrac{1}{4} \theta^{\a}{}_{\m} \theta^{\b}{}_{\n} w_{\l}{}^{\g} + \dfrac{1}{4} \theta^{\a}{}_{\m} \theta^{\b}{}_{\l} w_{\n}{}^{\g} -  \dfrac{1}{4} \theta^{\a}{}_{\l} \theta^{\b}{}_{\m} w_{\n}{}^{\g}
\eea
The second one is given by
\be A_{i 0 0}\ee  
\bea
(\mathcal{P}_{1}^{w})_{\l \m \n}^{\a \b \g}&=\dfrac{1}{12} \theta_{\n}{}^{\g} w^{\a}{}_{\m} w^{\b}{}_{\l} -  \dfrac{1}{6} \theta_{\m}{}^{\g} w^{\a}{}_{\n} w^{\b}{}_{\l} + \dfrac{1}{12} \theta_{\n}{}^{\g} w^{\a}{}_{\l} w^{\b}{}_{\m} + \dfrac{1}{12} \theta_{\l}{}^{\g} w^{\a}{}_{\n} w^{\b}{}_{\m} -\nonumber\\
&-  \dfrac{1}{6} \theta_{\m}{}^{\g} w^{\a}{}_{\l} w^{\b}{}_{\n} + \dfrac{1}{12} \theta_{\l}{}^{\g} w^{\a}{}_{\m} w^{\b}{}_{\n} + \dfrac{1}{12} \theta^{\b}{}_{\n} w^{\a}{}_{\m} w_{\l}{}^{\g} -  \dfrac{1}{6} \theta^{\b}{}_{\m} w^{\a}{}_{\n} w_{\l}{}^{\g} -\nonumber\\
&-  \dfrac{1}{6} \theta^{\a}{}_{\n} w^{\b}{}_{\m} w_{\l}{}^{\g} + \dfrac{1}{3} \theta^{\a}{}_{\m} w^{\b}{}_{\n} w_{\l}{}^{\g} + \dfrac{1}{12} \theta^{\b}{}_{\n} w^{\a}{}_{\l} w_{\m}{}^{\g} + \dfrac{1}{12} \theta^{\b}{}_{\l} w^{\a}{}_{\n} w_{\m}{}^{\g} - \nonumber\\
&- \dfrac{1}{6} \theta^{\a}{}_{\n} w^{\b}{}_{\l} w_{\m}{}^{\g} -  \dfrac{1}{6} \theta^{\a}{}_{\l} w^{\b}{}_{\n} w_{\m}{}^{\g} -  \dfrac{1}{6} \theta^{\b}{}_{\m} w^{\a}{}_{\l} w_{\n}{}^{\g} + \dfrac{1}{12} \theta^{\b}{}_{\l} w^{\a}{}_{\m} w_{\n}{}^{\g} +\nonumber\\
&+ \dfrac{1}{3} \theta^{\a}{}_{\m} w^{\b}{}_{\l} w_{\n}{}^{\g} - \dfrac{1}{6} \theta^{\a}{}_{\l} w^{\b}{}_{\m} w_{\n}{}^{\g}
\eea

And there is also a spin 1 trace given by
\bea
(\mathcal{P}_1^{t})_{\l \m \n}^{\a \b \g}&=- \dfrac{1}{6} \theta^{a}{}_{\n} \theta^{\b\g} \theta_{\l\m} + \dfrac{1}{12} \theta^{ag} \theta^{\b}{}_{\n} \theta_{\l\m} + \dfrac{1}{12} \theta^{\a\b} \theta^{\g}{}_{\n} \theta_{\l\m} + \dfrac{1}{3} \theta^{a}{}_{\m} \theta^{\b\g} \theta_{ln} -\nonumber\\
&  -\dfrac{1}{6} \theta^{ag} \theta^{\b}{}_{\m} \theta_{ln} -  \dfrac{1}{6} \theta^{\a\b} \theta^{\g}{}_{\m} \theta_{ln} -  \dfrac{1}{6} \theta^{a}{}_{\l} \theta^{\b\g} \theta_{\m\n} + \dfrac{1}{12} \theta^{ag} \theta^{\b}{}_{\l} \theta_{\m\n} + \dfrac{1}{12} \theta^{\a\b} \theta^{\g}{}_{\l} \theta_{\m\n}
\eea

\item Finally, there are two spin 2 projectors. The first one is the transverse traceless spin two component

\be
{1\over 2}\left(A_{j 0 i}+A_{i 0 j}\right)-{1\over 3} \d_{ij}  \sum_{k=1}^3 A_{k 0 k} 
\ee
with projector
\bea
(\mathcal{P}_2)_{\l \m \n}^{\a \b \g}&= - \dfrac{1}{6} \theta^{\b}{}_{\n} \theta^{\g}{}_{\m} w^{\a}{}_{\l} -  \dfrac{1}{6} \theta^{\b}{}_{\m} \theta^{\g}{}_{\n} w^{\a}{}_{\l} + \dfrac{1}{9} \theta^{\b\g} \theta_{\m\n} w^{\a}{}_{\l} + \dfrac{1}{3} \theta^{\b}{}_{\n} \theta^{\g}{}_{\l} w^{\a}{}_{\m} \nonumber\\
&+ \dfrac{1}{3} \theta^{\b}{}_{\l} \theta^{\g}{}_{\n} w^{\a}{}_{\m} -  \dfrac{2}{9} \theta^{\b\g} \theta_{\l\n} w^{\a}{}_{\m} -  \dfrac{1}{6} \theta^{\b}{}_{\m} \theta^{\g}{}_{\l} w^{\a}{}_{\n} -  \dfrac{1}{6} \theta^{\b}{}_{\l} \theta^{\g}{}_{\m} w^{\a}{}_{\n} \nonumber\\
&+ \dfrac{1}{9} \theta^{\b\g} \theta_{\l\m} w^{\a}{}_{\n} + \dfrac{1}{12} \theta^{\a}{}_{\n} \theta^{\g}{}_{\m} w^{\b}{}_{\l} + \dfrac{1}{12} \theta^{\a}{}_{\m} \theta^{\g}{}_{\n} w^{\b}{}_{\l} -  \dfrac{1}{18} \theta^{\a\g} \theta_{\m\n} w^{\b}{}_{\l} -\nonumber\\
&  -\dfrac{1}{6} \theta^{\a}{}_{\n} \theta^{\g}{}_{\l} w^{\b}{}_{\m} -  \dfrac{1}{6} \theta^{\a}{}_{\l} \theta^{\g}{}_{\n} w^{\b}{}_{\m} + \dfrac{1}{9} \theta^{\a\g} \theta_{\l\n} w^{\b}{}_{\m} + \dfrac{1}{12} \theta^{\a}{}_{\m} \theta^{\g}{}_{\l} w^{\b}{}_{\n} \nonumber\\
&+ \dfrac{1}{12} \theta^{\a}{}_{\l} \theta^{\g}{}_{\m} w^{\b}{}_{\n} -  \dfrac{1}{18} \theta^{\a\g} \theta_{\l\m} w^{\b}{}_{\n} +\dfrac{1}{12} \theta^{\a}{}_{\n} \theta^{\b}{}_{\m} w^{\g}{}_{\l} + \dfrac{1}{12} \theta^{\a}{}_{\m} \theta^{\b}{}_{\n} w^{\g}{}_{\l} \nonumber\\
&- \dfrac{1}{18} \theta^{\a\b} \theta_{\m\n} w^{\g}{}_{\l} -  \dfrac{1}{6} \theta^{\a}{}_{\n} \theta^{\b}{}_{\l} w^{\g}{}_{\m} -  \dfrac{1}{6} \theta^{\a}{}_{\l} \theta^{\b}{}_{\n} w^{\g}{}_{\m} 
+ \dfrac{1}{9} \theta^{\a\b} \theta_{ln} w^{\g}{}_{\m} \nonumber\\
&+ \dfrac{1}{12} \theta^{\a}{}_{\m} \theta^{\b}{}_{\l} w^{\g}{}_{\n} + \dfrac{1}{12} \theta^{\a}{}_{\l} \theta^{\b}{}_{\m} w^{\g}{}_{\n} -  \dfrac{1}{18} \theta^{\a\b} \theta_{\l\m} w^{\g}{}_{\n}
\eea

The second one corresponds to the spin 2 traceless connection field

\be
A^T_{ijk}\equiv A_{ijk}-{ 2 t^1_i-t^2_i\over 5}\d_{jk}-{3 t^2_j-t^1_j\over 10}\d_{ik}-{3 t^2_k-t^1_k\over 10}~\d_{ij}
\ee
with projector
\bea
(\mathcal{P}_{2}^{s})_{\l \m \n}^{\a \b \g}&=- \dfrac{1}{6} \theta^{a}{}_{\n} \theta^{\b}{}_{\m} \theta^{\g}{}_{\l} + \dfrac{1}{3} \theta^{a}{}_{\m} \theta^{\b}{}_{\n} \theta^{\g}{}_{\l} -  \dfrac{1}{6} \theta^{a}{}_{\n} \theta^{\b}{}_{\l} \theta^{\g}{}_{\m} -  \dfrac{1}{6} \theta^{a}{}_{\l} \theta^{\b}{}_{\n} \theta^{\g}{}_{\m} + \dfrac{1}{3} \theta^{a}{}_{\m} \theta^{\b}{}_{\l} \theta^{\g}{}_{\n} -\nonumber\\
& - \dfrac{1}{6} \theta^{a}{}_{\l} \theta^{\b}{}_{\m} \theta^{\g}{}_{\n} + \dfrac{1}{6} \theta^{a}{}_{\n} \theta^{\b\g} \theta_{\l\m} -  \dfrac{1}{12} \theta^{ag} \theta^{\b}{}_{\n} \theta_{\l\m} -  \dfrac{1}{12} \theta^{\a\b} \theta^{\g}{}_{\n} \theta_{\l\m} -  \dfrac{1}{3} \theta^{a}{}_{\m} \theta^{\b\g} \theta_{ln} +\nonumber\\
&+ \dfrac{1}{6} \theta^{ag} \theta^{\b}{}_{\m} \theta_{ln} + \dfrac{1}{6} \theta^{\a\b} \theta^{\g}{}_{\m} \theta_{ln} + \dfrac{1}{6} \theta^{a}{}_{\l} \theta^{\b\g} \theta_{\m\n} -  \dfrac{1}{12} \theta^{ag} \theta^{\b}{}_{\l} \theta_{\m\n} -  \dfrac{1}{12} \theta^{\a\b} \theta^{\g}{}_{\l} \theta_{\m\n}
\eea
\ei

Therefore, the spin content in this sector is
\be
\underline{20}_H= 2 \left(\underline{2}\right)\oplus 3\,\left( \underline{1}\right)\oplus \,\left( \underline{0}\right)
\ee

Finally, the closure relation in this space reads

\be \cP_0^s+\cP_1^s+\cP_1^w+\cP_1^t+\cP_2+\cP_2^s=\cP_H\label{closureH}\ee

\subsection{Mixed operators completing a basis of $\cL(\cA,\cA)$}\label{Bmix}
Let us represent by $\cL(\cA,\cA)$ the space of linear mappings from $\cA$ in $\cA$.
It is plain that a basis is given by (again, with implicit permutations)
\bea
&M_1\equiv k_\m k_\n k_\l k_\a k_\b k_\g &&M_2\equiv \eta_{\n\l}k_\m k_\a k_\b k_\g\nonumber\\
&M_3\equiv \eta_{\m\n} k_\l k_\a k_\b k_\g &&M_4\equiv \eta_{\m\a} k_\n k_\g k_\b k_\l \nonumber\\ 
&M_5\equiv \eta_{\m\b} k_\n k_\l k_\a k_\g &&M_6\equiv \eta_{\n\b}k_\m k_\l k_\a k_\g\nonumber\\
&M_7\equiv \eta_{\m\a} \eta_{\b\g} k_\n k_\l &&M_8\equiv \eta_{\m\b} \eta_{\a\g} k_\n k_\l\nonumber\\
&M_9\equiv \eta_{\a\b} \eta_{\l\g} k_\m k_\n &&M_{10}\equiv\eta_{\a\l} \eta_{\b\g} k_\m k_\n \nonumber\\
&M_{11}\equiv \eta_{\n\l}\eta_{\b\g} k_\m k_\a &&M_{12}\equiv \eta_{\n\b}\eta_{\l\g} k_\m k_\a\nonumber\\
&M_{13}\equiv \eta_{\n\l}\eta_{\a\g}k_\m k_\b &&M_{14}\equiv \eta_{\n\a}\eta_{\l\g} k_\m k_\b\nonumber\\
&M_{15}\equiv\eta_{\m\a}\eta_{\n\b}\eta_{\l\g} &&M_{16}\equiv\eta_{\m\a}\eta_{\n\l}\eta_{\b\g}\nonumber\\
&M_{17}\equiv\eta_{\m\b}\eta_{\n\a}\eta_{\l\g} &&M_{18}\equiv\eta_{\m\b}\eta_{\n\l}\eta_{\a\g}\nonumber\\ &M_{19}\equiv\eta_{\m\n}\eta_{\l\a}\eta_{\b\g} &&M_{20}\equiv\eta_{\m\n}\eta_{\l\b}\eta_{\a\g}\nonumber\\
&M_{21}\equiv\eta_{\m\l}\eta_{\n\a}\eta_{\b\g}&&M_{22}\equiv\eta_{\m\l}\eta_{\n\b}\eta_{\a\g}\nonumber\
\eea

So far, we have obtained 12 different operators that satisfy the closure relation.

Given the fact that we have obtained up to now 12 projectors, which added to the identity in our space --see \eqref{closureS} and \eqref{closureH}--, it is plain  that we are 10 operators short in order to get a complete basis on the space $\cL(\cA,\cA)$.  The remaining operators (which are not, in general, projectors) correspond to the mixing of equal spin components of $A$. In the same sense that $P_0^\times$ in \eqref{PX} corresponds to the mixing of the two spin 0 components of $h_{\m\n}$. Hence, we are going to classify them by their spin.
\par

\bi
\item There are three of them  with spin 0
\bea
(\mathscr{P}_0^{sw})^{\a \b \g \l \m \n} &= \dfrac{4}{9} \theta^{\mu \nu} \omega^{\alpha \lambda} \omega^{\beta \gamma} + \dfrac{1}{9} \theta^{\lambda \nu} \omega^{\alpha \mu} \omega^{\beta \gamma} + \dfrac{1}{9} \theta^{\lambda \mu} \omega^{\alpha \nu} \omega^{\beta \gamma} + \dfrac{1}{9} \theta^{\mu \nu} \omega^{\alpha \gamma} \omega^{\beta \lambda} \nonumber\\ \nonumber
&-  \dfrac{2}{9} \theta^{\lambda \nu} \omega^{\alpha \gamma} \omega^{\beta \mu} -  \dfrac{2}{9} \theta^{\lambda \mu} \omega^{\alpha \gamma} \omega^{\beta \nu} + \dfrac{1}{9} \theta^{\mu \nu} \omega^{\alpha \beta} \omega^{\gamma \lambda} -  \dfrac{2}{9} \theta^{\lambda \nu} \omega^{\alpha \beta} \omega^{\gamma \mu} \\ \nonumber
&-  \dfrac{2}{9} \theta^{\lambda \mu} \omega^{\alpha \beta} \omega^{\gamma \nu} + \dfrac{1}{9} \theta^{\beta \gamma} \omega^{\alpha \nu} \omega^{\lambda \mu} -  \dfrac{2}{9} \theta^{\alpha \gamma} \omega^{\beta \nu} \omega^{\lambda \mu} -  \dfrac{2}{9} \theta^{\alpha \beta} \omega^{\gamma \nu} \omega^{\lambda \mu} \\ \nonumber 
&+ \dfrac{1}{9} \theta^{\beta \gamma} \omega^{\alpha \mu} \omega^{\lambda \nu} -  \dfrac{2}{9} \theta^{\alpha \gamma} \omega^{\beta \mu} \omega^{\lambda \nu} -  \dfrac{2}{9} \theta^{\alpha \beta} \omega^{\gamma \mu} \omega^{\lambda \nu} + \dfrac{4}{9} \theta^{\beta \gamma} \omega^{\alpha \lambda} \omega^{\mu \nu} \\ 
&+ \dfrac{1}{9} \theta^{\alpha \gamma} \omega^{\beta \lambda} \omega^{\mu \nu} + \dfrac{1}{9} \theta^{\alpha \beta} \omega^{\gamma \lambda} \omega^{\mu \nu}
\eea

\bea
\nonumber(\mathscr{P}_0^{ws})^{\a \b \g \l \m \n} &= \dfrac{1}{9} \theta^{\mu \nu} \omega^{\alpha \lambda} \omega^{\beta \gamma} + \dfrac{1}{9} \theta^{\lambda \nu} \omega^{\alpha \mu} \omega^{\beta \gamma} + \dfrac{1}{9} \theta^{\lambda \mu} \omega^{\alpha \nu} \omega^{\beta \gamma} + \dfrac{1}{9} \theta^{\mu \nu} \omega^{\alpha \gamma} \omega^{\beta \lambda} \\ \nonumber 
&+ \dfrac{1}{9} \theta^{\lambda \nu} \omega^{\alpha \gamma} \omega^{\beta \mu} + \dfrac{1}{9} \theta^{\lambda \mu} \omega^{\alpha \gamma} \omega^{\beta \nu} + \dfrac{1}{9} \theta^{\mu \nu} \omega^{\alpha \beta} \omega^{\gamma \lambda} + \dfrac{1}{9} \theta^{\lambda \nu} \omega^{\alpha \beta} \omega^{\gamma \mu} \\ \nonumber
&+ \dfrac{1}{9} \theta^{\lambda \mu} \omega^{\alpha \beta} \omega^{\gamma \nu} + \dfrac{1}{9} \theta^{\beta \gamma} \omega^{\alpha \nu} \omega^{\lambda \mu} + \dfrac{1}{9} \theta^{\alpha \gamma} \omega^{\beta \nu} \omega^{\lambda \mu} + \dfrac{1}{9} \theta^{\alpha \beta} \omega^{\gamma \nu} \omega^{\lambda \mu} \\ \nonumber 
&+ \dfrac{1}{9} \theta^{\beta \gamma} \omega^{\alpha \mu} \omega^{\lambda \nu} + \dfrac{1}{9} \theta^{\alpha \gamma} \omega^{\beta \mu} \omega^{\lambda \nu} + \dfrac{1}{9} \theta^{\alpha \beta} \omega^{\gamma \mu} \omega^{\lambda \nu} + \dfrac{1}{9} \theta^{\beta \gamma} \omega^{\alpha \lambda} \omega^{\mu \nu} \\ 
&+ \dfrac{1}{9} \theta^{\alpha \gamma} \omega^{\beta \lambda} \omega^{\mu \nu} + \dfrac{1}{9} \theta^{\alpha \beta} \omega^{\gamma \lambda} \omega^{\mu \nu}
\eea

\bea
(\mathscr{P}_0^{x})^{\a \b \g \l \m \n} &= \dfrac{1}{6} \theta^{\alpha \gamma} \theta^{\lambda \nu} \omega^{\beta \mu} + \dfrac{1}{6} \theta^{\alpha \gamma} \theta^{\lambda \mu} \omega^{\beta \nu} + \dfrac{1}{6} \theta^{\alpha \beta} \theta^{\lambda \nu} \omega^{\gamma \mu} + \dfrac{1}{6} \theta^{\alpha \beta} \theta^{\lambda \mu} \omega^{\gamma \nu}
\eea

\item There are six with spin 1

\bea
\nonumber(\mathscr{P}_1^{wx})^{\a \b \g \l \m \n} &=  \dfrac{1}{4} \theta^{\g\n} \omega^{\a\m} \omega^{\b\l} + \dfrac{1}{4} \theta^{\g\m} \omega^{\a\n} \omega^{\b\l} + \dfrac{1}{4} \theta^{\g\n} \omega^{\a\l} \omega^{\b\m} + \dfrac{1}{4} \theta^{\g\m} \omega^{\a\l} \omega^{\b\n} \\  &+ \dfrac{1}{4} \theta^{\b\n} \omega^{\a\m} \omega^{\g\l} + \dfrac{1}{4} \theta^{\b\m} \omega^{\a\n} \omega^{\g\l} 
+ \dfrac{1}{4} \theta^{\b\n} \omega^{\a\l} \omega^{\g\m} + \dfrac{1}{4} \theta^{\b\m} \omega^{\a\l} \omega^{\g\n}
\eea

\bea
\nonumber(\mathscr{P}_1^{ws})^{\a \b \g \l \m \n} &= \dfrac{1}{9} \theta^{\g\n} \theta^{\l\m} \omega^{\a\b} + \dfrac{1}{9} \theta^{\g\m} \theta^{\l\n} \omega^{\a\b} + \dfrac{1}{9} \theta^{\g\l} \theta^{\m\n} \omega^{\a\b} + \dfrac{1}{9} \theta^{\b\n} \theta^{\l\m} \omega^{\a\g} \\ \nonumber
&+ \dfrac{1}{9} \theta^{\b\m} \theta^{\l\n} \omega^{\a\g} + \dfrac{1}{9} \theta^{\b\l} \theta^{\m\n} \omega^{\a\g} + \dfrac{1}{9} \theta^{\a\n} \theta^{\l\m} \omega^{\b\g} + \dfrac{1}{9} \theta^{\a\m} \theta^{\l\n} \omega^{\b\g} \\ \nonumber 
&+ \dfrac{1}{9} \theta^{\a\l} \theta^{\m\n} \omega^{\b\g} + \dfrac{1}{9} \theta^{\a\n} \theta^{\b\g} \omega^{\l\m} + \dfrac{1}{9} \theta^{\a\g} \theta^{\b\n} \omega^{\l\m} + \dfrac{1}{9} \theta^{\a\b} \theta^{\g\n} \omega^{\l\m} \\ \nonumber 
&+ \dfrac{1}{9} \theta^{\a\m} \theta^{\b\g} \omega^{\l\n} + \dfrac{1}{9} \theta^{\a\g} \theta^{\b\m} \omega^{\l\n} + \dfrac{1}{9} \theta^{\a\b} \theta^{\g\m} \omega^{\l\n} + \dfrac{1}{9} \theta^{\a\l} \theta^{\b\g} \omega^{\m\n} \\ 
&+ \dfrac{1}{9} \theta^{\a\g} \theta^{\b\l} \omega^{\m\n} + \dfrac{1}{9} \theta^{\a\b} \theta^{\g\l} \omega
\eea

\bea
\nonumber(\mathscr{P}_1^{sw})^{\a \b \g \l \m \n} &=  \dfrac{1}{9} \theta^{\g\n} \theta^{\l\m} \omega^{\a\b} + \dfrac{1}{9} \theta^{\g\m} \theta^{\l\n} \omega^{\a\b} -  \dfrac{2}{9} \theta^{\g\l} \theta^{\m\n} \omega^{\a\b} + \dfrac{1}{9} \theta^{\b\n} \theta^{\l\m} \omega^{\a\g} \\ \nonumber 
&+ \dfrac{1}{9} \theta^{\b\m} \theta^{\l\n} \omega^{\a\g} -  \dfrac{2}{9} \theta^{\b\l} \theta^{\m\n} \omega^{\a\g} -  \dfrac{2}{9} \theta^{\a\n} \theta^{\l\m} \omega^{\b\g} -  \dfrac{2}{9} \theta^{\a\m} \theta^{\l\n} \omega^{\b\g} \\ \nonumber 
&+ \dfrac{4}{9} \theta^{\a\l} \theta^{\m\n} \omega^{\b\g} -  \dfrac{2}{9} \theta^{\a\n} \theta^{\b\g} \omega^{\l\m} + \dfrac{1}{9} \theta^{\a\g} \theta^{\b\n} \omega^{\l\m} + \dfrac{1}{9} \theta^{\a\b} \theta^{\g\n} \omega^{\l\m} \\ \nonumber 
&-  \dfrac{2}{9} \theta^{\a\m} \theta^{\b\g} \omega^{\l\n} + \dfrac{1}{9} \theta^{\a\g} \theta^{\b\m} \omega^{\l\n} + \dfrac{1}{9} \theta^{\a\b} \theta^{\g\m} \omega^{\l\n} + \dfrac{4}{9} \theta^{\a\l} \theta^{\b\g} \omega^{\m\n} \\ 
&-  \dfrac{2}{9} \theta^{\a\g} \theta^{\b\l} \omega^{\m\n} -  \dfrac{2}{9} \theta^{\a\b} \theta^{\g\l} \omega^{\m\n}
\eea

\bea
\nonumber(\mathscr{P}_1^{sx})^{\a \b \g \l \m \n} &= - \dfrac{2}{9} \theta^{\g\n} \theta^{\l\m} \omega^{\a\b} -  \dfrac{2}{9} \theta^{\g\m} \theta^{\l\n} \omega^{\a\b} + \dfrac{1}{9} \theta^{\g\l} \theta^{\m\n} \omega^{\a\b} -  \dfrac{2}{9} \theta^{\b\n} \theta^{\l\m} \omega^{\a\g} \\ \nonumber 
&-  \dfrac{2}{9} \theta^{\b\m} \theta^{\l\n} \omega^{\a\g} + \dfrac{1}{9} \theta^{\b\l} \theta^{\m\n} \omega^{\a\g} + \dfrac{1}{9} \theta^{\a\n} \theta^{\l\m} \omega^{\b\g} + \dfrac{1}{9} \theta^{\a\m} \theta^{\l\n} \omega^{\b\g}  \\ \nonumber 
&+ \dfrac{4}{9} \theta^{\a\l} \theta^{\m\n} \omega^{\b\g} + \dfrac{1}{9} \theta^{\a\n} \theta^{\b\g} \omega^{\l\m} -  \dfrac{2}{9} \theta^{\a\g} \theta^{\b\n} \omega^{\l\m} -  \dfrac{2}{9} \theta^{\a\b} \theta^{\g\n} \omega^{\l\m}  \\ \nonumber
&+ \dfrac{1}{9} \theta^{\a\m} \theta^{\b\g} \omega^{\l\n} -  \dfrac{2}{9} \theta^{\a\g} \theta^{\b\m} \omega^{ln} -  \dfrac{2}{9} \theta^{\a\b} \theta^{\g\m} \omega^{\l\n} + \dfrac{4}{9} \theta^{\a\l} \theta^{\b\g} \omega^{\m\n}  \\ 
&+ \dfrac{1}{9} \theta^{\a\g} \theta^{\b\l} \omega^{\m\n} + \dfrac{1}{9} \theta^{\a\b} \theta^{\g\l} \omega^{\m\n}
\eea

\bea
\nonumber(\mathscr{P}_1^{ss})^{\a \b \g \l \m \n} &= \dfrac{1}{18} \theta^{\alpha \nu} \theta^{\beta \gamma} \theta^{\lambda \mu} + \dfrac{1}{72} \theta^{\alpha \gamma} \theta^{\beta \nu} \theta^{\lambda \mu} + \dfrac{1}{72} \theta^{\alpha \beta} \theta^{\gamma \nu} \theta^{\lambda \mu} + \dfrac{1}{18} \theta^{\alpha \mu} \theta^{\beta \gamma} \theta^{\lambda \nu} \nonumber\\ 
& + \dfrac{1}{72} \theta^{\alpha \gamma} \theta^{\beta \mu} \theta^{\lambda \nu} + \dfrac{1}{72} \theta^{\alpha \beta} \theta^{\gamma \mu} \theta^{\lambda \nu} + \dfrac{2}{9} \theta^{\alpha \lambda} \theta^{\beta \gamma} \theta^{\mu \nu} + \dfrac{1}{18} \theta^{\alpha \gamma} \theta^{\beta \lambda} \theta^{\mu \nu} \nonumber\\ 
&+ \dfrac{1}{18} \theta^{\alpha \beta} \theta^{\gamma \lambda} \theta^{\mu \nu}
\eea

\bea
\nonumber(\mathscr{P}_1^{wst})^{\a \b \g \l \m \n} &= - \dfrac{1}{18} \theta^{\gamma \nu} \theta^{\lambda \mu} \omega^{\alpha \beta} -  \dfrac{1}{18} \theta^{\gamma \mu} \theta^{\lambda \nu} \omega^{\alpha \beta} -  \dfrac{2}{9} \theta^{\gamma \lambda} \theta^{\mu \nu} \omega^{\alpha \beta} -  \dfrac{1}{18} \theta^{\beta \nu} \theta^{\lambda \mu} \omega^{\alpha \gamma} \\ \nonumber 
&-  \dfrac{1}{18} \theta^{\beta \mu} \theta^{\lambda \nu} \omega^{\alpha \gamma} -  \dfrac{2}{9} \theta^{\beta \lambda} \theta^{\mu \nu} \omega^{\alpha \gamma} + \dfrac{5}{18} \theta^{\alpha \nu} \theta^{\lambda \mu} \omega^{\beta \gamma} + \dfrac{5}{18} \theta^{\alpha \mu} \theta^{\lambda \nu} \omega^{\beta \gamma} \\ \nonumber
&+ \dfrac{1}{9} \theta^{\alpha \lambda} \theta^{\mu \nu} \omega^{\beta \gamma} -  \dfrac{2}{9} \theta^{\alpha \nu} \theta^{\beta \gamma} \omega^{\lambda \mu} -  \dfrac{1}{18} \theta^{\alpha \gamma} \theta^{\beta \nu} \omega^{\lambda \mu} -  \dfrac{1}{18} \theta^{\alpha \beta} \theta^{\gamma \nu} \omega^{\lambda \mu} \\ \nonumber
&-  \dfrac{2}{9} \theta^{\alpha \mu} \theta^{\beta \gamma} \omega^{\lambda \nu} -  \dfrac{1}{18} \theta^{\alpha \gamma} \theta^{\beta \mu} \omega^{\lambda \nu} -  \dfrac{1}{18} \theta^{\alpha \beta} \theta^{\gamma \mu} \omega^{\lambda \nu} + \dfrac{1}{9} \theta^{\alpha \lambda} \theta^{\beta \gamma} \omega^{\mu \nu} \\ 
&+ \dfrac{5}{18} \theta^{\alpha \gamma} \theta^{\beta \lambda} \omega^{\mu \nu} + \dfrac{5}{18} \theta^{\alpha \beta} \theta^{\gamma \lambda} \omega^{\mu \nu}
\eea

\item Finally, there is one more with spin 2
\bea
\nonumber(\mathscr{P}_2^{x})^{\a \b \g \l \m \n} &= \dfrac{1}{4} \theta^{\alpha \nu} \theta^{\gamma \lambda} \omega^{\beta \mu} + \dfrac{1}{4} \theta^{\alpha \lambda} \theta^{\gamma \nu} \omega^{\beta \mu} -  \dfrac{1}{6} \theta^{\alpha \gamma} \theta^{\lambda \nu} \omega^{\beta \mu} + \dfrac{1}{4} \theta^{\alpha \mu} \theta^{\gamma \lambda} \omega^{\beta \nu} \\ \nonumber
&+ \dfrac{1}{4} \theta^{\alpha \lambda} \theta^{\gamma \mu} \omega^{\beta \nu} -  \dfrac{1}{6} \theta^{\alpha \gamma} \theta^{\lambda \mu} \omega^{\beta \nu} + \dfrac{1}{4} \theta^{\alpha \nu} \theta^{\beta \lambda} \omega^{\gamma \mu} + \dfrac{1}{4} \theta^{\alpha \lambda} \theta^{\beta \nu} \omega^{\gamma \mu} \\ 
&-  \dfrac{1}{6} \theta^{\alpha \beta} \theta^{\lambda \nu} \omega^{\gamma \mu} + \dfrac{1}{4} \theta^{\alpha \mu} \theta^{\beta \lambda} \omega^{\gamma \nu} + \dfrac{1}{4} \theta^{\alpha \lambda} \theta^{\beta \mu} \omega^{\gamma \nu} -  \dfrac{1}{6} \theta^{\alpha \beta} \theta^{\lambda \mu} \omega^{\gamma \nu}
\eea
\ei
\newpage

\section{Spin content of the  antisymmetric connection field}\label{C}

In this appendix, we decompose the operators mediating between two connection fields $A_{\m\b\g}\equiv g_{\a\m}\Gamma^\a_{\b\g}$ -- antisymmetric in the las two indices because we consider torsionful connections which fulfill the metricity condition-- in terms of the spin projectors of this field. The procedure is analogue to the one followed in Appendices \ref{A} and \ref{B}.

The subspace ${\cal A}$ corresponds, in terms of representations of the tangent group SO(4), to the sum of a totally antisymmetric three-index tensor plus a tensor with the {\em hook} symmetry
\be
\{0,2\}\otimes \{1\}=\{0,3\}\oplus \{2,1\}
\ee

In terms of dimensions this is $\underline{24}=\underline{4}+\underline{20}$

\subsection{The totally antisymmetric tensor}\label{Basym}
We want to determine the spin content of the totally antisymmetric piece $A_{[\a\b\g]}$, in this case there are only two monomials we can form
\bea
&M_{23}=\d^{[a}_{[\l}\d^\b_\m\d^{\g]}_{\n]}\nonumber\\
&M_{24}=\d^{[a}_{[\l}\d^\b_\m k^{\g]}k_{\n]}\eea

The totally antisymmetric piece is represented as 
\be
\{0,3\}\quad\quad \young(~,~,~)
\ee
and the corresponding Young projectors are
\bea
&\left(\bar{P}_{\tiny{\young(\a,\b,\g)}}\right)_{\m\n\l}^{\a\b\g}\equiv {1\over 6}\bigg\{\d^\a_\m \d^\b_\n \d^\g_\l+\d^\b_\m\d^\g_\n\d^\a_\l+ \d^\g_\m\d^\a_\n\d^\b_\l- \d^\a_\m\d^\g_\n\d^\b_\l- \d^\b_\m\d^\a_\n\d^\g_\l-\d^\g_\m\d^\b_\n\d^\a_\l\bigg\}
=\nonumber\\
&={1\over 6}\left(1,1,1,-1,-1,-1\right)
\eea
where the notation of the projectors in the same as in Appendix \ref{B}.

We can decompose it in its spin componets as
\bi
\item First the spin 1 component
\be
{1\over 2}\left(A_{j 0 i}-A_{i 0 j}\right)
\ee 
with projector
\bea(\bar{P}_1)^{\a \b \g \l \m \n} &=-\frac{1}{6} \theta^{\a\n} \theta^{\b\m} \theta^{\g\l}+\frac{1}{6} \theta^{\a\m}\theta^{\b\n} \theta^{\g\l}+\frac{1}{6} \theta^{\a\n} \theta^{\b\l} \theta^{\g\m}-\nonumber\\&-\frac{1}{6} \theta^{\a\l} \theta^{\b\n} \theta^{\g\m}-\frac{1}{6} \theta^{\a\m} \theta^{\b\l} \theta^{\g\n}+\frac{1}{6} \theta^{\a\l} \theta^{\b\m} \theta^{\g\n}\eea
\item The spin 0 component is
\be
A_{[ijk]}
\ee 
with projector
\bea(\bar{P}_0)^{\a \b \g \l \m \n} &=-\frac{1}{6} \w^{\a\l} \theta^{\b\n}\theta^{\g\m}+\frac{1}{6} \w^{\a\l} \theta^{\b\m} \theta^{\g\n}+\frac{1}{6} \w^{\a\m} \theta^{\b\n} \theta^{\g\l}-\frac{1}{6} \w^{\a\m} \theta^{\b\l} \theta^{\g\n}-\nonumber\\&-\frac{1}{6} \w^{\a\n} \theta^{\b\m} \theta^{\g\l}+\frac{1}{6} \w^{\a\n} \theta(b,l) \theta^{\g\m}+\frac{1}{6} \theta^{\a\n} \w^{\b\l} \theta^{\g\m}-\frac{1}{6} \theta^{\a\m} \w^{\b\l} \theta^{\g\n}-\nonumber\\&-\frac{1}{6} \theta^{\a\n} \w^{\b\m} \theta^{\g\l}+\frac{1}{6} \theta^{\a\l} \w^{\b\m} \theta^{\g\n}+\frac{1}{6} \theta^{\a\m}\w^{\b\n} \theta^{\g\l}-\frac{1}{6} \theta^{\a\l} \w^{\b\n} \theta^{\g\m}-\nonumber\\&-\frac{1}{6} \theta^{\a\n}\theta^{\b\m}\w^{\g\l}+\frac{1}{6} \theta^{\a\m} \theta^{\b\n} \w^{\g\l}+\frac{1}{6} \theta^{\a\n} \theta^{\b\l} \w^{\g\m}-\frac{1}{6} \theta^{\a\l} \theta^{\b\n} \w^{\g\m}-\nonumber\\&-\frac{1}{6} \theta^{\a\m} \theta^{\b\l} \w^{\g\n}+\frac{1}{6} \theta^{\a\l} \theta^{\b\m}\w^{\g\n}\eea
\ei

Finally it is easy to check that
\bea
\left(\bar{P}\right)_{\m\n\l}^{\a\b\g}=(\bar{P}_1)_{\m\n\l}^{\a\b\g}+(\bar{P}_0)_{\m\n\l}^{\a\b\g}
\eea
In terms of dimensions this is $\underline{4}=(\underline{1})\oplus(\underline{0})$. 
\subsection{The antisymmetric Hook sector}\label{Bahook}
We determine the spin content of the antisymmetric hook piece $A_{\a[\b\g]}$, in this case there are six monomials
\bea
&M_{25}=\d^{\a}_{\l}\d^{[\b}_{[\m}\d^{\g]}_{\n]}\nonumber\\
&M_{26}=k^{\a}k_{\l}\,\d^{[\b}_{[\m}\d^{\g]}_{\n]}\nonumber\\
&M_{27}=k^{\a}\d^{[\b}_{\l}\, k_{[\m}\d^{\g]}_{\n]}\nonumber\\
&M_{28}=\d^{\a}_{[\m}k^{[\b}k_{\n]}\d^{\g]}_{\l}\nonumber\\
&M_{29}=\d^{\a}_{\l}\,k^{[\b} k_{[\m}\d^{\g]}_{\n]}\nonumber\\
&M_{30}=k^{\a}k_{\l}\,k^{[\b} k_{[\m}\d^{\g]}_{\n]}\nonumber\\
\eea

The antisymmetric hook part corresponds to the piece
\be
\{2,1\}\quad\quad \yng(2,1)
\ee

The Young projectors reads
\bea
\bar{\cP}_H\equiv \left(P_{\tiny{\young(\a \b,\g)}}\right)_{\m\n\l}^{\a\b\g}&\equiv {1\over 3}\bigg\{\d^\a_\m\d^\b_\n \d^\g_\l-\d^\a_\m\d_\n^\g\d^\b_\l +\dfrac{1}{2}\d^\a_\n\d^\b_\m\d^\g_\l-\dfrac{1}{2}\d^\a_\n\d^\b_\l\d^\g_\m+\dfrac{1}{2}\d^\a_\l\d^\b_\n\d^\g_\m-\dfrac{1}{2}\d^\a_\l\d^\b_\m\d^\g_\n\bigg\}=\nonumber\\
&={1\over 3}\left(1,\dfrac{1}{2},-\dfrac{1}{2},-1,\dfrac{1}{2},-\dfrac{1}{2}\right)
\eea

We can decompose it in its spin componets as
\bi
\item There are two  spin 2 component. The first one is the transverse traceless spin two component

\be
{1\over 2}\left(A_{j 0 i}+A_{i 0 j}\right)-{1\over 3} \d_{ij}  \sum_{k=1}^3 A_{k 0 k} 
\ee
with projector
\bea(\bar{\cP}_2)^{\a \b \g \l \m \n} &=\frac{1}{4} \theta^{\a\n} \w^{\b\m} \theta^{\g\l}+\frac{1}{4} \theta^{\a\l}\w^{\b\m} \theta^{\g\n}-\frac{1}{6} \theta^{\a\g} \w^{\b\m}\theta^{\n\l}-\frac{1}{4} \theta^{\a\m} \w^{\b\n} \theta^{\g\l}-\nonumber\\&-\frac{1}{4} \theta^{\a\l} \w^{\b\n} \theta^{\g\m}+\frac{1}{6} \theta^{\a\g} \w^{\b\n} \theta^{\m\l}-\frac{1}{4} \theta^{\a\n}\theta^{\b\l} \w^{\g\m}-\frac{1}{4} \theta^{\a\l} \theta^{\b\n} \w^{\g\m}+\nonumber\\&+\frac{1}{6} \theta^{\a\b} \w^{\g\m} \theta^{\n\l}+\frac{1}{4} \theta^{\a\m}\theta^{\b\l} \w^{\g\n}+\frac{1}{4} \theta^{\a\l} \theta^{\b\m} \w^{\g\n}-\frac{1}{6} \theta^{\a\b} \w^{\g\n} \theta^{\m\l}\eea

The second one corresponds to the spin 2 traceless connection field
\be
A^T_{ijk}\equiv A_{ijk}-\frac{1}{2}t_j\d_{ik}+\frac{1}{2}t_k\d_{ij}
\ee
where $t_i=\sum_{j=1}^3A_{jij}$, with projector
\bea(\bar{\cP}^s_2)^{\a \b \g \l \m \n} &=
\frac{1}{6} \theta^{\a\n} \theta^{\b\m} \theta^{\g\l}-\frac{1}{6} \theta^{\a\m} \theta^{\b\n} \theta^{\g\l}-\frac{1}{6} \theta^{\a\n} \theta^{\b\l} \theta^{\g\m}-\nonumber\\&-\frac{1}{3} \theta^{\a\l} \theta^{\b\n} \theta^{\g\m}+\frac{1}{6} \theta^{\a\m}\theta^{\b\l} \theta^{\g\n}+\frac{1}{3} \theta^{\a\l} \theta^{\b\m} \theta^{\g\n}+\nonumber\\&+\frac{1}{4} \theta^{\a\g} \theta^{\b\n} \theta^{\l\m}-\frac{1}{4} \theta^{\a\b} \theta^{\g\n} \theta^{\l\m}-\frac{1}{4} \theta^{\a\g} \theta{\b\m} \theta^{\l\n}+\frac{1}{4} \theta^{\a\b} \theta^{\g\m} \theta^{\l\n}\eea

\item There are three spin 1 components. First
\be
{1\over 2}\left(A_{j 0 i}-A_{i 0 j}\right)
\ee 
with projector
\bea(\bar{\cP}^s_1)^{\a \b \g \l \m \n}&=-\frac{1}{3} \w^{\a\l} \theta^{\b\n} \theta^{\g\m}+\frac{1}{3} \w^{\a\l} \theta^{\b\m} \theta^{\g\n}-\frac{1}{6} \w^{\a\m} \theta^{\b\n} \theta^{\g\l}+\nonumber\\&+\frac{1}{6} \w^{\a\m} \theta^{\b\l} \theta^{\g\n}+\frac{1}{6} \w^{\a\n} \theta^{\b\m} \theta^{\g\l}-\frac{1}{6} \w^{\a\n} \theta^{\b\l} \theta^{\g\m}-\frac{1}{6} \theta^{\a\n} \w^{\b\l} \theta^{\g\m}+\nonumber\\&+\frac{1}{6} \theta^{\a\m} \w^{\b\l} \theta^{\g\n}-\frac{1}{12} \theta^{\a\n} \w^{\b\m} \theta^{\g\l} +\frac{1}{12} \theta^{\a\l}  \w^{\b\m} \theta^{\g\n}+\nonumber\\&+\frac{1}{12} \theta^{\a\m} \w^{\b\n} \theta^{\g\l}-\frac{1}{12} \theta^{\a\l}  \w^{\b\n} \theta^{\g\m}+\frac{1}{6} \theta^{\a\n} \theta^{\b\m} \w^{\g\l}-\frac{1}{6} \theta^{\a\m}\theta^{\b\n} \w^{\g\l}+\nonumber\\&+\frac{1}{12} \theta^{\a\n} \theta^{\b\l}  \w^{\g\m}-\frac{1}{12} \theta^{\a\l}  \theta^{\b\n} \w^{\g\m}-\frac{1}{12} \theta^{\a\m} \theta^{\b\l} \w^{\g\n}+\frac{1}{12} \theta^{\a\l}  \theta^{\b\m} \w^{\g\n}\eea 

The second one is given by
\be A_{0i0}\ee
corresponding to
\bea(\bar{\cP}^w_1)^{\a \b \g \l \m \n} &=\frac{1}{2} \w^{\a\b} \theta^{\g\n} \w^{\l\m}-\frac{1}{2} w^{\a\g}\theta^{\b\n} \w^{\l\m}-\frac{1}{2} \w^{\a\b} \theta^{\g\m} \w^{\l\n}+\frac{1}{2} \w^{\a\g} \theta^{\b\m}\w^{\l\n}\eea 
And there is also a spin 1 trace 
\be \sum_{j=1}^3A_{jij}\ee
given by
\bea(\bar{\cP}^t_1)^{\a \b \g \l \m \n} &=-\frac{1}{4} \theta^{\a\g}\theta^{\b\n} \theta^{\l\m}+\frac{1}{4} \theta^{\a\b}\theta^{\g\n} \theta^{\l\m}+\frac{1}{4} \theta^{\a\g}\theta^{\b\m} \theta^{\l\n}-\frac{1}{4} \theta^{\a\b} \theta^{\g\m} \theta^{\l\n}\eea 
\item There is only one spin zero, a trace that is given by
\be 
\sum_{i=1}^3 A_{i 0 i}
\ee
that is
\bea(\bar{\cP}_0)^{\a \b \g \l \m \n} &=\frac{1}{6} \theta^{\a\g} \w^{\b\m} \theta^{\l\n}-\frac{1}{6} \theta^{\a\g} \w^{\b\n} \theta^{\l\m}-\frac{1}{6} \theta^{\a\b} \w^{\g\m}\theta^{\l\n}+\frac{1}{6} \theta^{\a\b} \w^{\g\n}\theta^{\l\m}\eea 
\ei
Finally it is easy to check that
\bea
\left(\bar{\cP}_{H}\right)_{\m\n\l}^{\a\b\g}=(\bar{\cP}_2)_{\m\n\l}^{\a\b\g}+(\bar{\cP}^s_2)_{\m\n\l}^{\a\b\g}+(\bar{\cP}^s_1)_{\m\n\l}^{\a\b\g}+(\bar{\cP}^w_1)_{\m\n\l}^{\a\b\g}+(\bar{\cP}^t_1)_{\m\n\l}^{\a\b\g}+(\bar{\cP}_0)_{\m\n\l}^{\a\b\g}
\eea
In terms of dimensions this is $\underline{20}=2(\underline{2})\oplus\,3(\underline{1})\oplus(\underline{0})$.
\par
These projectors agree with the ones obtained by  Sezgin and van Nieuwenhuizen in \cite{Sezgin:1979zf}.

\newpage

\section{Zero modes for $R^2$}\label{D}
In section 5 we had determined the quadratic one loop operator in the particular case where the lagrangian is proportional to $R^2$, the square of the scalar curvature.  
\bea
(K_{ R^2+\text{gf}})^{\m\n~\r\s}_{~\t~~\l} &= \dfrac{1}{\chi}\Big( \text{P}_0^w + 3 \ \text{P}_{0}^s + (3 - 9 \chi ) \ \mathcal{P}_0^{s} - 3 \ \mathscr{P}_0^{x} + \mathscr{P}_0^{sw}+ \mathscr{P}_0^{ws} + \text{P}_1^w - \dfrac{5}{3} \ \text{P}_{1}^s \nonumber \\
&+ \mathcal{P}_{1}^{w} + \dfrac{2}{3} \ \mathcal{P}_1^{t} - \mathscr{P}_1^{wx}+\mathscr{P}_1^{ws} + \mathscr{P}_1^{sw} + \mathscr{P}_1^{sx} + 4 \ \mathscr{P}_1^{ss} \Big)^{\m\n~\r\s}_{~\t~~\l} \ \Box 
\eea

It can be checked that this operator has 13 independent zero modes, which are written in terms of the spin operators acting on an arbitrary field $\Omega_{\a\b\g}\in {\cal A}$ as
\bea
&Z_1\equiv \left(\text{P}_0^w + \text{P}_{0}^s - \mathscr{P}_0^{ws}\right)_{\l \m \n}^{\a \b \g} \ \Omega_{\a\b\g} \nonumber \\
&Z_2\equiv \left(-\text{P}_1^w + \text{P}_{1}^s + 3 \mathcal{P}_{1}^{w} - \tfrac{3}{8} \mathscr{P}_1^{sw} - \tfrac{3}{2} \mathscr{P}_1^{wst} \right)_{\l \m \n}^{\a \b \g} \ \Omega_{\a\b\g}  \nonumber \\
&Z_3\equiv \left( 2 \mathcal{P}_{1}^{w} + \mathcal{P}_{1}^{t} - \tfrac{3}{2} \mathscr{P}_1^{sw} \right)_{\l \m \n}^{\a \b \g} \ \Omega_{\a\b\g} \nonumber \\
&Z_4\equiv \left(-2 \text{P}_1^w + \mathcal{P}_{1}^{w} + \mathscr{P}_1^{ws} - \tfrac{1}{8}\mathscr{P}_1^{sw}  - \tfrac{1}{2} \mathscr{P}_1^{wst} \right)_{\l \m \n}^{\a \b \g} \ \Omega_{\a\b\g} \nonumber \\
&Z_5\equiv \left(-2 \text{P}_1^w + \mathcal{P}_{1}^{w}  - \tfrac{3}{4}\mathscr{P}_1^{sw}  + \mathscr{P}_1^{sx} - \mathscr{P}_1^{wst} \right)_{\l \m \n}^{\a \b \g} \ \Omega_{\a\b\g}  \nonumber \\
&Z_6\equiv\left(- \tfrac 7 6  \text{P}_1^w +  \tfrac{14}{3} \mathcal{P}_{1}^{w}  - \tfrac{21}{16}\mathscr{P}_1^{ws} + \mathscr{P}_1^{ss} - \tfrac 7 4 \mathscr{P}_1^{wst}  \right)_{\l \m \n}^{\a \b \g} \ \Omega_{\a\b\g} \nonumber \\
& Z_7\equiv(\mathcal{P}_1^{s})_{\l \m \n}^{\a \b \g} \ \Omega_{\a\b\g} \nonumber \\
& Z_8\equiv(\mathscr{P}_1^{wx})_{\l \m \n}^{\a \b \g} \ \Omega_{\a\b\g} \nonumber \\
&Z_9\equiv (\text{P}_2)_{\l \m \n}^{\a \b \g} \ \Omega_{\a\b\g} \nonumber \\
&Z_{10}\equiv (\mathcal{P}_{2})_{\l \m \n}^{\a \b \g} \ \Omega_{\a\b\g} \nonumber \\
&Z_{11}\equiv (\mathcal{P}_{2}^{s})_{\l \m \n}^{\a \b \g} \ \Omega_{\a\b\g} \nonumber \\
&Z_{12}\equiv (\mathscr{P}_2^\text{x})_{\l \m \n}^{\a \b \g} \ \Omega_{\a \b \g} \nonumber \\
&Z_{13}\equiv(\text{P}_3)_{\l \m \n}^{\a \b \g} \ \Omega_{\a \b \g}
\eea

It is quite remarkable that the system has extra  gauge symmetries at one loop order that are not present in the exact lagrangian. The physical meaning of this is discussed in the main body of the paper.
\newpage

\section{Fun with $S_3$}\label{E}
Let us highlight the procedure to get the spin projectors in a systematic way. Denoting the elements of permutation group of three elements $S_3$ acting on $T_{\a\b\g}\in T\times T\times T$ as
\bea
&g_1\equiv \d^\a_\m \d^\b_\n \d^\g_\l\nonumber\\
&g_2\equiv \d^\b_\m \d^\g_\n \d^\a_\l\nonumber\\
&g_3\equiv \d^\g_\m \d^\a_\n \d^\b_\l\nonumber\\
&g_4\equiv \d^\a_\m \d^\g_\n \d^\b_\l\nonumber\\
&g_5\equiv \d^\b_\m \d^\a_\n \d^\g_\l\nonumber\\
&g_6\equiv \d^\g_\m \d^\b_\n \d^\a_\l
\eea
The most general projector in this space can be written as
\be
P\equiv \sum_{i=1}^{i=6} C_i\, g_i\equiv \begin{pmatrix}U\\V\end{pmatrix}
\ee
where we have defined the column vectors 
\bea
&U\equiv \begin{pmatrix}C_1\\C_2\\C_3\end{pmatrix}\nonumber\\
&V\equiv \begin{pmatrix}C_4\\C_5\\C_6\end{pmatrix}
\eea
Those operators are not symmetric ones; rather the transpose operator is given by 
\be
\left(C_1,C_2,C_3,C_4,C_5,C_6\right)^T=\left(C_1,C_3,C_2,C_4,C_5,C_6\right)
\ee
It is important to keep this in mind when multiplying projectors.
\par
On the other hand, it is not difficult to establish that
\bea
\left(P^{\prime\prime}\right)_{\vec{\m}}^{\vec{a}}\equiv \sum_{\vec{c}}P_{\vec{c}}^{\vec{a}}.\left(P^\prime\right)_{\vec{\m}}^{\vec{c}}= M \begin{pmatrix}U^\prime\\V^\prime\end{pmatrix}\equiv \begin{pmatrix}U^{\prime\prime}\\V^{\prime\prime}\end{pmatrix}=\begin{pmatrix} A U^\prime+B V^\prime\\
B U^\prime+ A V^\prime\end{pmatrix}
\eea
with
\bea
&M\equiv \begin{pmatrix}A&B\\B&A\end{pmatrix}\nonumber\\
&A\equiv \begin{pmatrix}C_1&C_3 &C_2\\C_2&C_1&C_3\\C_3&C_2&C_1\end{pmatrix}\nonumber\\
&B\equiv \begin{pmatrix}C_4&C_5&C_6\\C_5&C_6&C_4\\C_6&C_4&C_5\end{pmatrix}
\eea
All this implies that
\be
\left[P,P^\prime\right]=\begin{pmatrix}0\\C_{54}+C_{65}+C_{46}\\C_{64}+C_{45}+C_{56}\\C_{52}+C_{63}+C_{35}+C_{28}\\C_{52}+C_{63}+C_{35}+C_{26}\\C_{62}+C_{43}+C_{24}+C_{36}\\C_{42}+C_{53}+C_{34}+C_{25}\end{pmatrix}
\ee
where
\be
C_{ab}\equiv C_a C^\prime_b-C_b C^\prime_a
\ee
These formulas make it trivial to check all assertions about projectors, which have been nevertheless also verified  with xAct \cite{Martingarcia}.


\begin{thebibliography}{99}

\bibitem{Stelle} 
K.~S.~Stelle,
``Renormalization of Higher Derivative Quantum Gravity,''
Phys.\ Rev.\ D {\bf 16}, 953 (1977).
doi:10.1103/PhysRevD.16.953\\
``Classical Gravity with Higher Derivatives,''
Gen.\ Rel.\ Grav.\  {\bf 9} (1978) 353.
doi:10.1007/BF00760427

\bibitem{Alvarez}
E.~Alvarez,
``Quantum Gravity: A Pedagogical Introduction To Some Recent Results,''
Rev.\ Mod.\ Phys.\  {\bf 61} (1989) 561.
doi:10.1103/RevModPhys.61.561

\bibitem{AlvarezGaume}
L.~Alvarez-Gaume, A.~Kehagias, C.~Kounnas, D.~L\"ust and A.~Riotto,
``Aspects of Quadratic Gravity,''
Fortsch.\ Phys.\  {\bf 64} (2016) no.2-3,  176
doi:10.1002/prop.201500100
[arXiv:1505.07657 [hep-th]]

\bibitem{AAG}
E.~Alvarez, J.~Anero and S.~Gonzalez-Martin,
``Quadratic gravity in first order formalism,''
JCAP {\bf 1710} (2017) no.10,  008
doi:10.1088/1475-7516/2017/10/008
[arXiv:1703.07993 [hep-th]].\\
E.~Alvarez and S.~Gonzalez-Martin,
``Weyl Gravity Revisited,''
JCAP {\bf 1702} (2017) no.02,  011
doi:10.1088/1475-7516/2017/02/011
[arXiv:1610.03539 [hep-th]].\\
E.~Alvarez, J.~Anero, S.~Gonzalez-Martin and R.~Santos-Garcia,
``A candidate for an UV completion: quadratic gravity in first order formalism,''
arXiv:1710.01764 [hep-th].

\bibitem{Einhorn:2017icw}
M.~B.~Einhorn and D.~R.~T.~Jones,
``Renormalizable, asymptotically free gravity without ghosts or tachyons,''
Phys.\ Rev.\ D {\bf 96} (2017) no.12,  124025
doi:10.1103/PhysRevD.96.124025
[arXiv:1710.03795 [hep-th]].

\bibitem{Donoghue:2018izj} 
J.~F.~Donoghue and G.~Menezes,
``Gauge Assisted Quadratic Gravity: A Framework for UV Complete Quantum Gravity,''
Phys.\ Rev.\ D {\bf 97}, no. 12, 126005 (2018)
doi:10.1103/PhysRevD.97.126005
[arXiv:1804.04980 [hep-th]].
\bibitem{Salvio:2017qkx} 

A.~Salvio and A.~Strumia,
``Agravity up to infinite energy,''
Eur.\ Phys.\ J.\ C {\bf 78}, no. 2, 124 (2018)
doi:10.1140/epjc/s10052-018-5588-4
[arXiv:1705.03896 [hep-th]].

\bibitem{Ehlers}
J. Ehlers, F. A. E. Pirani and A. Schild, 
"The geometry of free fall and light propagation"
in: General Relativity, papers in honour of J. L. Synge. Edited by L. O'Raifeartaigh. Oxford, Clarendon Press 1972, pp. 63-84.

\bibitem{Ferreira:2018itt}
P.~G.~Ferreira, C.~T.~Hill and G.~G.~Ross,
``Inertial Spontaneous Symmetry Breaking and Quantum Scale Invariance,''
arXiv:1801.07676 [hep-th].

\bibitem{Einhorn:2015lzy} 
M.~B.~Einhorn and D.~R.~T.~Jones,
``Induced Gravity I: Real Scalar Field,''
JHEP {\bf 1601}, 019 (2016)
doi:10.1007/JHEP01(2016)019
[arXiv:1511.01481 [hep-th]].

\bibitem{Salvio:2014soa}
A.~Salvio and A.~Strumia,
``Agravity,''
JHEP {\bf 1406} (2014) 080
doi:10.1007/JHEP06(2014)080
[arXiv:1403.4226 [hep-ph]].

\bibitem{Shapiro:1994yt} 
I.~L.~Shapiro,
Mod.\ Phys.\ Lett.\ A {\bf 9}, 1985 (1994)
doi:10.1142/S0217732394001842
[hep-th/9403077].
S.~D. Odintsov I.~L.~Shapiro,
``Curvature phase transition in $R^2$ quantum gravity and induction of Einstein gravity,''
Theor. \ Math. \ Phys. , {\bf 90}, 319 (1992)
doi:10.1142/S0217732394001842
[hep-th/9403077].

\bibitem{Barnes}
K. J. Barnes,
 Unpublished (part of Ph. D. Thesis at University of London)
(1963)\\
R. J. Rivers,
"Lagrangian Theory ~or Neutral Massive Spin-2 Fields",
 Nuov. Cim. 34 (1964) 386.

\bibitem{VanNieuwenhuizen:1973fi} 
P.~Van Nieuwenhuizen,
``On ghost-free tensor lagrangians and linearized gravitation,''
Nucl.\ Phys.\ B {\bf 60}, 478 (1973).
doi:10.1016/0550-3213(73)90194-6

\bibitem{Creminelli}
  P.~Creminelli, A.~Nicolis, M.~Papucci and E.~Trincherini,
  ``Ghosts in massive gravity,''
  JHEP {\bf 0509} (2005) 003
  doi:10.1088/1126-6708/2005/09/003
  [hep-th/0505147].
  %
\bibitem{Bekaert}
  X.~Bekaert, N.~Boulanger and P.~Sundell,
  ``How higher-spin gravity surpasses the spin two barrier: no-go theorems versus yes-go examples,''
  Rev.\ Mod.\ Phys.\  {\bf 84} (2012) 987
  doi:10.1103/RevModPhys.84.987
  [arXiv:1007.0435 [hep-th]].






\bibitem{Martingarcia}
J.~ M.~ Martin-Garcia et.al. xAct: Efficient tensor computer algebra for Mathematica.2002-2013.url:http://xact.es/




\bibitem{Sezgin:1979zf}
E.~Sezgin and P.~van Nieuwenhuizen,
Phys.\ Rev.\ D {\bf 21} (1980) 3269.
doi:10.1103/PhysRevD.21.3269
		
\end{thebibliography}
\end{document}